\documentclass[twocolumn,prd]{revtex4-1}
\usepackage{color}
\usepackage{amsmath}
\usepackage{braket}
\usepackage[T1]{fontenc}
\usepackage{url}
\usepackage[english]{babel}
\usepackage{graphicx}
\usepackage{subfigure}
\usepackage{graphicx}% Include figure files
\usepackage{dcolumn}% Align table columns on decimal point
\usepackage{bm}% bold math
\usepackage{subfig}
\usepackage{float}
\usepackage{url}

\usepackage{grffile}

\newcommand{\Real}{\mbox{\rm Re\,}}
\newcommand{\Tr}{\mbox{\rm Tr}}

\newcommand{\be}{\begin{equation}}
\newcommand{\ee}{\end{equation}}
\newcommand{\bea}{\begin{eqnarray}}
\newcommand{\eea}{\end{eqnarray}}
\newcommand{\non}{\nonumber}
\newcommand{\bit}{\begin{itemize}}
\newcommand{\eit}{\end{itemize}}

\usepackage{verbatim}

\graphicspath{{figs/}}

\begin{document}

\title{Gluon screening mass at finite temperature from Landau gauge gluon propagator in lattice QCD }

\author{P. J. Silva}
\email{psilva@teor.fis.uc.pt}
\author{O. Oliveira}
\email{orlando@fis.uc.pt}
\affiliation{CFC, Departamento de F\'{\i}sica, Faculdade de Ci\^encias e Tecnologia, Universidade de Coimbra, 3004-516 Coimbra, Portugal}
\author{P. Bicudo}
\email{bicudo@ist.utl.pt}
\author{N. Cardoso}
\email{nunocardoso@cftp.ist.utl.pt; Address after September 2013: NCSA, University of Illinois, Urbana IL 61801, USA}
\affiliation{CFTP, Departamento de F\'{\i}sica, Instituto Superior T\'{e}cnico,
Av. Rovisco Pais, 1049-001 Lisboa, Portugal}

%%%%%%%%%%%%%%%%%%%%%%%%%    abstract   %%%%%%%%%%%%%
%.......................................................................................................................................
%.......................................................................................................................................
\begin{abstract}
We address the interpretation of the Landau gauge gluon propagator at finite temperature as a massive type bosonic propagator.  Using pure gauge SU(3) lattice simulations at a fixed lattice volume $\sim(6.5fm)^3$, we compute the electric and magnetic form factors, extract a gluon mass from Yukawa-like fits, and study its temperature dependence. This is relevant both for the Debye screening at high temperature $T$ and for confinement at low $T$.
\end{abstract}

\pacs{11.15.Ha,11.10.Wx,14.70.Dj}

\maketitle

%%%%%%%%%%%%%%%%%%%%%%%%%    Introduction   %%%%%%%%%%%
%.......................................................................................................................................
%.......................................................................................................................................
\section{Introduction}

Finite temperature $SU(3)$ pure Yang-Mills theory has a first order transition at the critical temperature
$T = T_c \sim 270$ MeV~\cite{Fingberg:1992ju,Lucini:2003zr}. 
For temperatures below $T_c$ gluons are confined, i.e. do not behave as free particles.
On the other hand, for  $T > T_c$ gluons become deconfined and, for 
sufficiently high $T$, they can be treated perturbatively~\cite{Smilga:1996cm,Andersen:2004fp}.
The order parameter for the confinement-deconfinement phase transition is the Polyakov loop $P_L$, 
a path order product of exponentials of the gluon field time-like component. 
For $T< T_c$ the center symmetry of gauge group is unbroken and $ P_L = 0$. On the other hand, for
$T > T_c$ the center symmetry is spontaneously broken and $P_L \neq 0 $. 
This change in the behavior of the Polyakov loop with temperature is rooted in the gluodynamics and 
should also be seen in the correlation functions of the gluon fields, such as the gluon propagator.

At zero temperature, the momentum space gluon propagator in the Landau gauge is described
by a single scalar function $D(p^2)$. For $T \ne 0$ the propagator requires two scalar functions 
$D_T(p^2;T)$ and $D_L(p^2;T)$ that are associated with the gluon transverse and longitudinal degrees of 
freedom, respectively. 
Recently, a number of lattice QCD simulations dedicated to the study of the two point gluon correlation
function
(see, for example,~\cite{Cucchieri:2007ta,Fischer:2010fx,Bornyakov:2010nc,Cucchieri:2011ga,
Bornyakov:2011jm,Aouane:2011fv,Cucchieri:2011di,Maas:2011ez,Cucchieri:2012nx,Maas:2011se,
Cucchieri:2012gb,Oliveira:2012hx,Oliveira:2012uy,Silva:2013ak} and references therein) have shown that
$D_L(p^2;T)$ and $D_T(p^2;T)$ are non trivial functions of $T$ and that their nature changes as $T$ crosses
$T_c$. Indeed, in~\cite{Aouane:2011fv,Maas:2011ez} the authors identify possible order parameters for 
the confinement-deconfinement phase transition directly related with the gluon 
propagator.

The lattice QCD simulations show that the electric $D_L (p^2;T)$ and magnetic  $D_T (p^2;T)$ form factors
are finite for all momenta and temperatures. 
An interpretation of the gluon propagator form factors in terms of quasi-particle massive bosons is welcome 
when building effective models to describe the hadronic phase diagram   ~\cite{Costa:2010pp,Bicudo:2012wt}
and also provides a check to 
the temperature dependent perturbative approach to QCD
~\cite{Shuryak:1977ut,Gross:1980br,Weiss:1981ev}.

A gluon screening mass can be identified with the pole of the gluon two point correlation function 
in momentum space or, in real space, with the exponential decay of the propagator at large distances. 
Although the gluon propagator is not gauge invariant, it has been conjectured that the 
gluon screening mass associated with the longitudinal degrees of freedom, i.e. the Debye gluon mass, 
is independent of the gauge~\cite{Toimela:1982ht,Furusawa:1983gb,Braaten:1991gm,Shrauner:1992gh,
Rebhan:1993uz,Amemiya:1998jz,Aguilar:2002tc,Dudal:2003by,Kondo:2006sa,Kondo:2006ih,Dudal:2006xd,
Ford:2009qh,Cucchieri:2011ig,Gongyo:2012jb}. 
At finite temperature, the Debye mass is related to the screening of the color interaction
which is expected to occur at high $T$~\cite{Shuryak:1977ut,Gross:1980br,Weiss:1981ev}. 
Note that at zero temperature, a finite and non-vanishing gluon mass is linked with 
gluon confinement~\cite{Cornwall:1981zr,Aguilar:2008xm,Oliveira:2010xc}.

Above the deconfinement temperature of QCD, finite temperature field theory, see 
e.g.~\cite{Smilga:1996cm,Andersen:2004fp,Kapusta:2006pm}, 
treats the gluons as an ideal gas of massive particles. The electric propagator acquires a pole which defines the 
gluon Debye mass $m_D \sim g T$, where $g$ is the strong coupling constant. 
For example, $m_D$ is used in Boltzmann models for heavy ion interactions~\cite{Xu:2008av}.
Similarly, for the transverse degrees of freedom one can define a magnetic mass $\sim g^2 T$, 
see e.g.~\cite{Kapusta:2006pm}.

The Debye mass can be computed in $SU(N)$ using perturbation theory and, to next to leading 
order in the strong coupling constant $g$, is given by~\cite{Arnold:1995bh}
\be
{ m_D \over  g T} =   {N \over 3}   +  g \left[ \alpha  + {N  \over 4 \pi}   \log \left( N \over 3 g  \right)  \right] +{ \cal O} ( g^2)   \ ,
\label{Eq:perturbative}
\ee
where $\alpha$ is a non-perturbative constant, in the sense that it is not computable within perturbation theory. 
For very high temperatures and up to $T \sim 10^4 T_c$, $m_D$ was measured using lattice simulations for the 
gauge group SU(2)~\cite{Heller:1997nqa}. 
The simulations show $m_D \propto T$ and an $\alpha$, associated with the slope of 
$m_D(T)$, that is large for all $T$'s. 
Moreover, the static quark-antiquark free energy has been computed at
finite $T$ using lattice QCD and the mass dimension screening parameter of the quark-antiquark potential
is compatible with a linear $T$ behavior~\cite{Nakamura:2004wra,Petreczky:2004pz,Karsch:2004ik,
Kaczmarek:2005zp,Kaczmarek:2005gi,Kaczmarek:2005ui,Hubner:2007qh,Doring:2007uh,Bicudo:2008gs,
Bicudo:2009pb}.

For large $T$, the color interaction is suppressed due to a Debye mass $m_D \ne 0$.
However, near the critical temperature $T_c$ it is not clear what value it should take~\cite{Greiner:1996wv}. 
Pure gauge QCD has a first order phase transition at $T_c = 270$ MeV and this suggests a vanishing gluon 
mass at $T = T_c$.  However, recently~\cite{Bicudo:2012wt}, using the experimental data for heavy ions 
and, in particular, the kaon to pion multiplicity ratio, a finite gluon $m_g = 0.32 \pm 0.07$ GeV was 
estimated at $T_c$.

On the other hand, the analogy between QCD and superconductors leads to a dual massive gluon at low 
temperature $T\sim 0$. In superconductors, the screening of the magnetic field in the London equation 
has a direct relation with an effective mass, the inverse of the magnetic penetration length, of the interaction 
particle, i. e. the photon. The dual gluon mass was studied using lattice QCD techniques
in~\cite{Burdanov:2000rw,Jia:2005sp,Suzuki:2004uz,Suganuma:2004gq,Suganuma:2004ij,
Suganuma:2003ds,Kumar:2004fj}; see, e.g.~\cite{Cardoso:2010kw} for a review on the dual gluon and 
gluon effective masses computed using several non-perturbative approaches. 
Recently,  the penetration length started to be computed with gauge invariant lattice QCD 
techniques~\cite{Cardoso:2010kw,Cardaci:2010tb,Cea:2012qw}. 
If some constituent gluon models assume a vanishing gluon 
mass~\cite{Mathieu:2005wc,Mathieu:2006bp,Mathieu:2008bf,Mathieu:2008pb,Boulanger:2008aj,
Mathieu:2008me,Mathieu:2008up,Buisseret:2009yv,Mathieu:2009sg,Mathieu:2011zzb} at $T\sim 0$, others 
consider a constituent gluon with a finite  
mass~\cite{Szczepaniak:1995cw,LlanesEstrada:2000jw,LlanesEstrada:2005jf,Bicudo:2006sd,
Hou:2001ig,Hou:2002jv}.
For an infrared gluon mass in the range 0.5 to 1.0 GeV, 
the constituent massive gluon models are also consistent with the large glueball masses as
predicted by lattice 
QCD~\cite{Ishikawa:1982ax,Ishikawa:1982tb,Ishikawa:1982bk,Ishikawa:1982uf,
Ishikawa:1983xg,Ishikawa:1983js,Schierholz:1983ir,deForcrand:1984rs,deForcrand:1985ww,
Carpenter:1987em,Teper:1987wt,Michael:1987wf,Teper:1987ws,
Michael:1988be,Michael:1988jr,Teper:1998te,Stephenson:1989pu,Bitar:1991wr,Kogut:1991ts,Moretto:1993dc,
Johnson:2000qz,Hart:2001fp,Meyer:2002mk,Meyer:2003wx,Lucini:2004my,Meyer:2004jc,
Morningstar:1997ff,Morningstar:1999rf,Chen:2005mg}.

A non-vanishing gluon mass is clearly a non-perturbative feature of QCD. For example, at zero temperature
and in perturbation theory, the gluon mass $m_g$, taken as a pole in the propagator, vanishes to all orders.
At zero temperature, the decoupling solution of the Dyson-Schwinger 
equations~\cite{Cornwall:1981zr,Aguilar:2011xe} generates a running effective gluon mass $m_g(p^2)$. 
Furthermore, lattice QCD results~\cite{Oliveira:2010xc} are compatible with the same type of running mass.

In this paper, we compute gluon masses $m_g(T)$, associated with the longitudinal and transverse gluon propagator 
form factors, for various temperatures and up to $2 T_c$ using lattice QCD simulations.
The gluon mass is identified by modeling the lattice data 
~\cite{Heller:1997nqa, Nakamura:2003pu, Maas:2011ez}
and we consider two different definitions: (i) an infrared
pole mass assuming a Yukawa functional form; (ii) a generalization of the functional form which reproduces the
zero temperature lattice and Dyson-Schwinger propagators~\cite{Cornwall:1981zr,Aguilar:2011xe,Oliveira:2010xc}.
Furthermore, we also investigate the mass scale associated with the zero momentum form factors. 
As discussed in~\cite{Maas:2011ez}, we confirm that this mass scale can also be used as an order parameter 
for the confinement-deconfinement phase transition.

Ideally, one would like to access the full complex $p^2$ plane to determine the poles of the propagator.
This is a nontrivial problem that would require, for example, the computation of the spectral 
density as discussed in~\cite{Oliveira:2012eu,Dudal:2013vf,Dudal:2013yva}.
The analytic structure of the gluon propagator has also been investigated within the Dyson-Schwinger approach 
in~\cite{Karplus:1958zz,Landau:1959fi,Cutkosky:1960sp,Fischer:2009jm,Dudal:2010wn,Windisch:2012zd,
Windisch:2012sz,Strauss:2012dg,Windisch:2013mg,Windisch:2013dxa}. However, given the approximations 
involved in the calculation and the dificulty of numerical computation, the outcome of the Dyson-Schwinger 
equations requires an independent confirmation.

The paper is organized as follows. 
In Section \ref{SecII} we detail the lattice QCD setup, the
gauge fixing procedure, the computation of the gluon propagator at finite $T$, the renormalization of the lattice
data and comment on the systematics.
In section \ref{resultados_finite_T} we summarize our results for the longitudinal and transverse components 
of the gluon propagator.
In Section \ref{calculo_gluon_mass} we detail our results for the gluon mass as a function of the temperature,
taking the various definitions referred in the previous paragraph. Finally in Section \ref{conclusoes} we resume and
conclude.

%.......................................................................................................................................
%.......................................................................................................................................
%.......................................................................................................................................
\section{Lattice setup, gauge fixing and the gluon propagator \label{SecII}}

%+++++++++++++++++++++++++++++++++++++++++++++++++++++++++++++++++++++++++++
%+++++++++++++++++++++++++++++++++++++++++++++++++++++++++++++++++++++++++++
%+++++++++++++++++++++++++++++++++++++++++++++++++++++++++++++++++++++++++++
\begin{table}[t!]
\begin{center}
\begin{tabular}{c@{\hspace{0.5cm}}c@{\hspace{0.3cm}}c@{\hspace{0.3cm}}r@{\hspace{0.4cm}}l@{\hspace{0.3cm}}l}
\hline
Temp.            & $\beta$ & $L_s$ &  \multicolumn{1}{c}{$L_t$} & \multicolumn{1}{c}{$a$} & \multicolumn{1}{c}{$1/a$} \\
 (MeV)           &              &            &            & \multicolumn{1}{c}{(fm)} & \multicolumn{1}{c}{(GeV)} \\
\hline
121 &   6.0000 & 64 & 16 & 	0.1016 &  	1.9426 \\
162 &   6.0000 & 64 & 12 & 	0.1016 & 	1.9426 \\
194 &   6.0000 & 64 & 10 & 	0.1016 & 	1.9426 \\
243 &   6.0000 & 64 &   8  & 	0.1016 & 	1.9426 \\
260 &   6.0347 & 68 &   8  & 	0.09502 & 	2.0767 \\
265 &   5.8876 & 52 &   6  & 	0.1243 & 	1.5881 \\
275 &   6.0684 & 72 &   8  & 	0.08974 & 	2.1989 \\
285 &   5.9266 & 56 &   6 & 	0.1154 & 	1.7103 \\
290 &   6.1009 & 76 &   8 & 	0.08502 & 	2.3211 \\
305 &   6.1326 & 80 &   8 & 	0.08077 & 	2.4432 \\
324 &   6.0000 & 64 &   6 & 	0.1016	 &      1.9426 \\
366 &   6.0684 & 72 &   6 & 	0.08974	 &      2.1989 \\
397 &   5.8876 & 52 &   4 & 	0.1243	 &      1.5881 \\
428 &   5.9266 & 56 &   4 & 	0.1154	 &      1.7103 \\
458 &   5.9640 & 60 &   4 & 	0.1077	 &      1.8324 \\
486 &   6.0000 & 64 & 	4 & 	0.1016	 &      1.9426 \\
\hline
\end{tabular}
\end{center}
\caption{Lattice setup used for the computation of the gluon propagator at finite temperature. The $\beta$ was adjusted to have $L_s \, a \simeq 6.5$ fm.
              See text for details.}
\label{tempsetup}
\end{table}

The lattice simulations reported here have been performed on a lattice $L_s^3 \times L_t$ 
using the Wilson gauge action for the gauge group SU(3) and for various $\beta$.
The temperature is taken as the inverse of the lattice time extension $T = 1 / L_t$ in physical units.
Note that in this work we consider a constant physical volume $\sim (6.5$ fm$)^3$. 
Given that the typical scale for color interaction is $\sim 1$ fm, 
the finite volume effects are expected to be small.

%+++++++++++++++++++++++++++++++++++++++++++++++++++++++++++++++++++++++++++
%+++++++++++++++++++++++++++++++++++++++++++++++++++++++++++++++++++++++++++
%+++++++++++++++++++++++++++++++++++++++++++++++++++++++++++++++++++++++++++
\begin{figure}[t!] %  figure placement: here, top, bottom, or page
\vspace{0.5cm}
\begin{center}
\includegraphics[width=1.0\columnwidth]{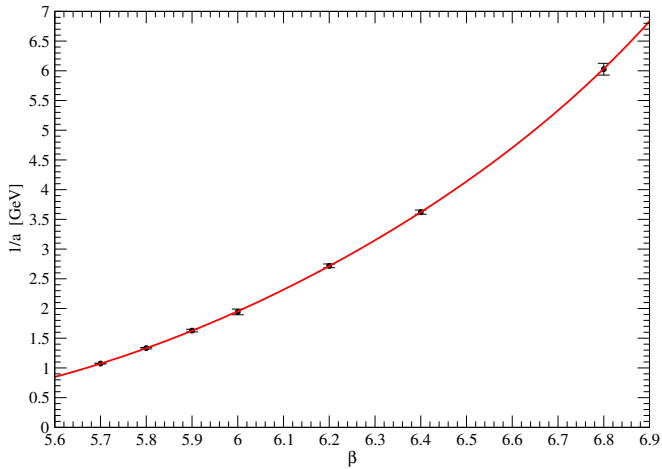}
\end{center}
\caption{$1/a(\beta)$ data from~\cite{Bali:1992ru} and the fit to Eq. (\ref{Eq:a_beta}).}
\label{fig:a_beta}
\end{figure}

For converting the simulation results into physical units we rely
on~\cite{Bali:1992ru}, where the lattice spacing was set from
the string tension for a number of $\beta$ values. In order to be able to keep the spatial volume at 
$\sim (6.5$ fm$)^3$ and access various $T$, the results of~\cite{Bali:1992ru} were fitted to the 
functional form considered in~\cite{Necco:2001xg}, i.e. to
\begin{equation}
  \frac{1}{a( \beta )} =    \exp\left\{ b_0  +  b_1 (\beta - 6)  + b_2  (\beta - 6)^2 + b_3  (\beta-6)^3 \right\} \ .
   \label{Eq:a_beta}
\end{equation}
The fit, see Fig. \ref{fig:a_beta}, gives $b_0 = 0.6702(87)$, $b_1 = 1.768(21)$, $b_2 = -0.68(10)$ and 
$b_3 = 0.29(14)$ for a $\chi^2/d.o.f. = 0.027$. In the following, we will always use the relation $a(\beta)$ 
as given by Eq. (\ref{Eq:a_beta}).

The lattice setup considered in the current work is described in Tab. \ref{tempsetup}. 
The generation of the gauge configurations was done using the Chroma library \cite{Edwards:2004sx}. 

%.......................................................................................................................................
%.......................................................................................................................................
%.......................................................................................................................................
\subsection{Landau gauge fixing}

In lattice QCD, the fundamental fields are the link variables $U_\mu (x) \in $ SU(3), which are related to the  gluon fields $A^a_\mu$ by 
\begin{equation}
  U_\mu (x) ~ = ~ \exp \Big\{ i \, a \, g_0 \, A_\mu (x + a \hat{e}_\mu / 2) \Big\} \ ;
\end{equation}
$\hat{e}_\mu$ are unit vectors along the $\mu$ direction. In QCD the fields related by gauge transformations
\bea
  U_\mu (x) & \longrightarrow & g(x) ~ U_\mu (x) ~ 
 g^\dagger (x + a \hat{e}_\mu) \, ,
\\ \non
 g &\in& SU(3) \, ,
\eea
are physically equivalent and they define the gauge orbits. For the computation of the QCD Green's functions it is 
enough to consider one field from each orbit. The choice of a single configuration in the gauge orbits is 
known as gauge fixing.

In this work, we consider the minimal Landau gauge which, on the lattice, means maximizing, for each gauge 
configuration and on its orbit, the functional
\begin{equation}
	F_U[g]=\frac{1}{4 N_cV} \sum_x \sum_\mu \Real \left[ \Tr\left(  g(x)U_\mu(x)g^\dagger(x+\mu)  \right) \right] \, ,
\end{equation}
where $N_c$ the dimension of the gauge group and $V$ the lattice volume.
It can be shown, see e.g.~\cite{Oliveira:2003wa}, that picking a maximum of $F_U[g]$ on a gauge orbit is equivalent to the usual 
continuum Landau gauge condition 
\be
\partial_\mu A^a_\mu = 0
\ee
and implies also the positiveness of the Faddeev-Popov determinant. 

In this work, the functional $F_U[g]$ was maximized using the Fourier accelerated steepest descent 
method as defined in~\cite{Davies:1987vs}. 
The evolution and convergence of the gauge fixing process was monitored by
\begin{equation}
	\theta = \frac{1}{N_c V}\sum_x \Tr\left[\Delta(x)\Delta^\dagger(x)\right]
\end{equation}
where
\begin{equation}
	\Delta(x) = \sum_\nu\left[ U_\nu(x-a\hat{\nu})-U_\nu(x) - \text{h.c.} - \text{trace} \right].
\end{equation}
%and
%\begin{equation}
%	\Delta_{-\nu}\left(U_\mu(x)\right) = U_\mu(x-a\hat{\nu})-U_\mu(x) \ .
%\end{equation}
The function $\Delta(x)$ is the lattice version of $\partial_\mu A_\mu$. The number $\theta$ gives
the mean value of $\partial_\mu A_\mu=0$ evaluated over all space-time lattice points per color degree of freedom.
In all the results shown below, gauge fixing was stopped only when $\theta \le 10^{-15}$. Gauge fixing was
implemented using Chroma \cite{Edwards:2004sx} and PFFT \cite{Pippig:2011} libraries.

 %+++++++++++++++++++++++++++++++++++++++++++++++++++++++++++++++++++++++++++
%+++++++++++++++++++++++++++++++++++++++++++++++++++++++++++++++++++++++++++
%+++++++++++++++++++++++++++++++++++++++++++++++++++++++++++++++++++++++++++
\vspace{-0.1cm}
\begin{figure*}[th!] %  figure placement: here, top, bottom, or page
   \begin{center}
   \subfigure{ \includegraphics[width=0.85\columnwidth]{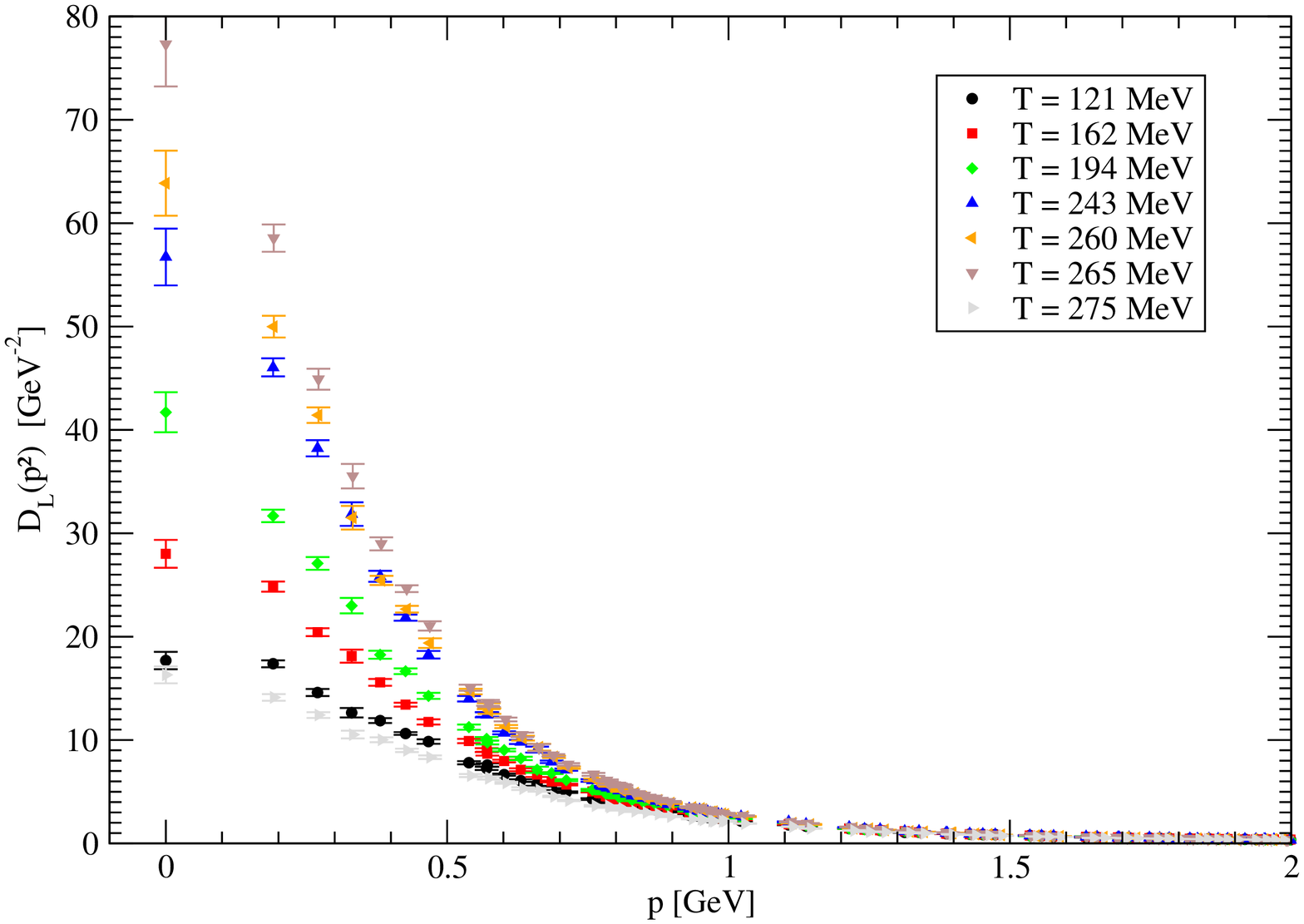} } \quad
   \subfigure{ \includegraphics[width=0.85\columnwidth]{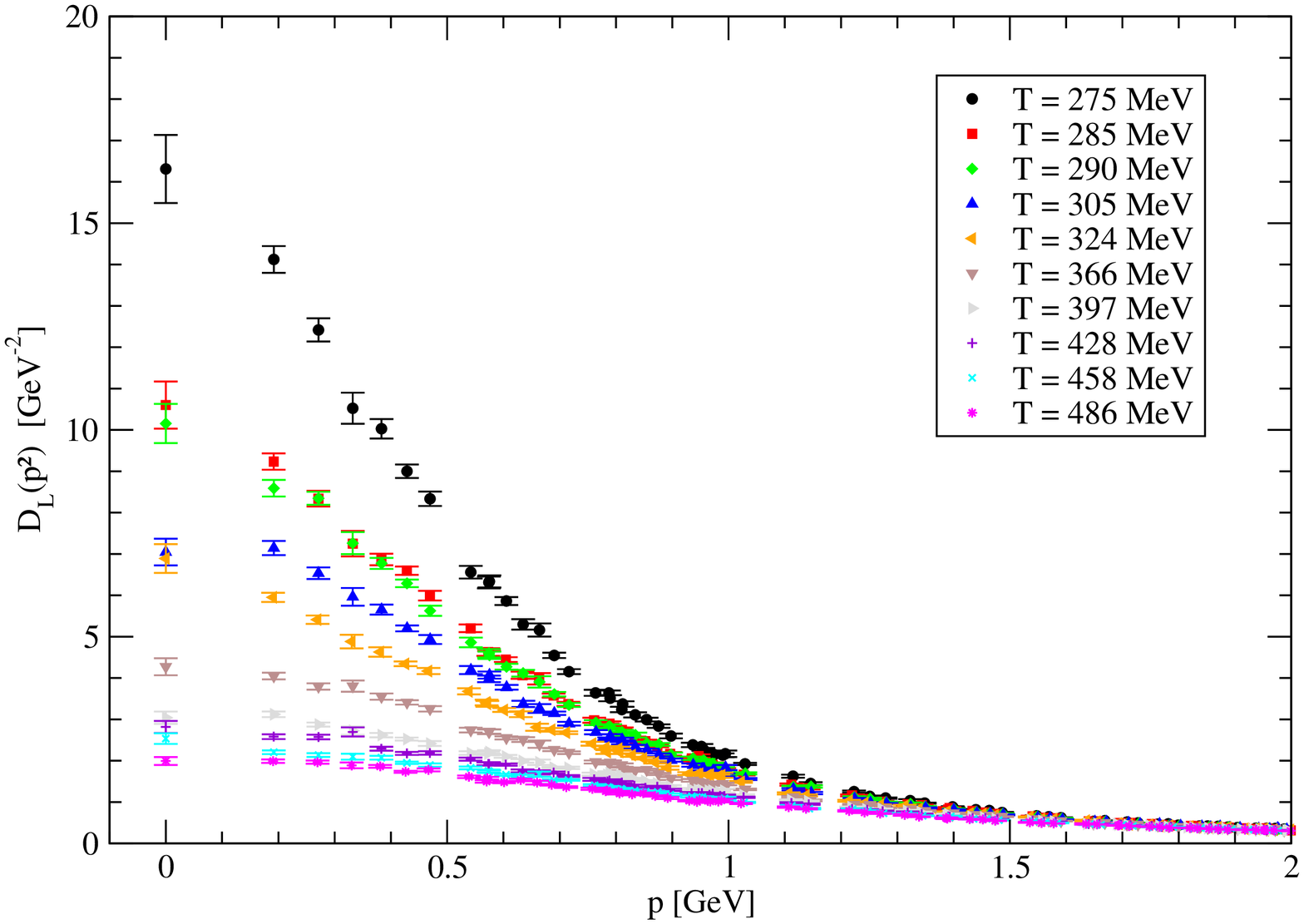} } \\
   \vspace{0.7cm}
   \subfigure{ \includegraphics[width=0.85\columnwidth]{gluon_Trans_up275MeV.eps} } \quad
   \subfigure{ \includegraphics[width=0.85\columnwidth]{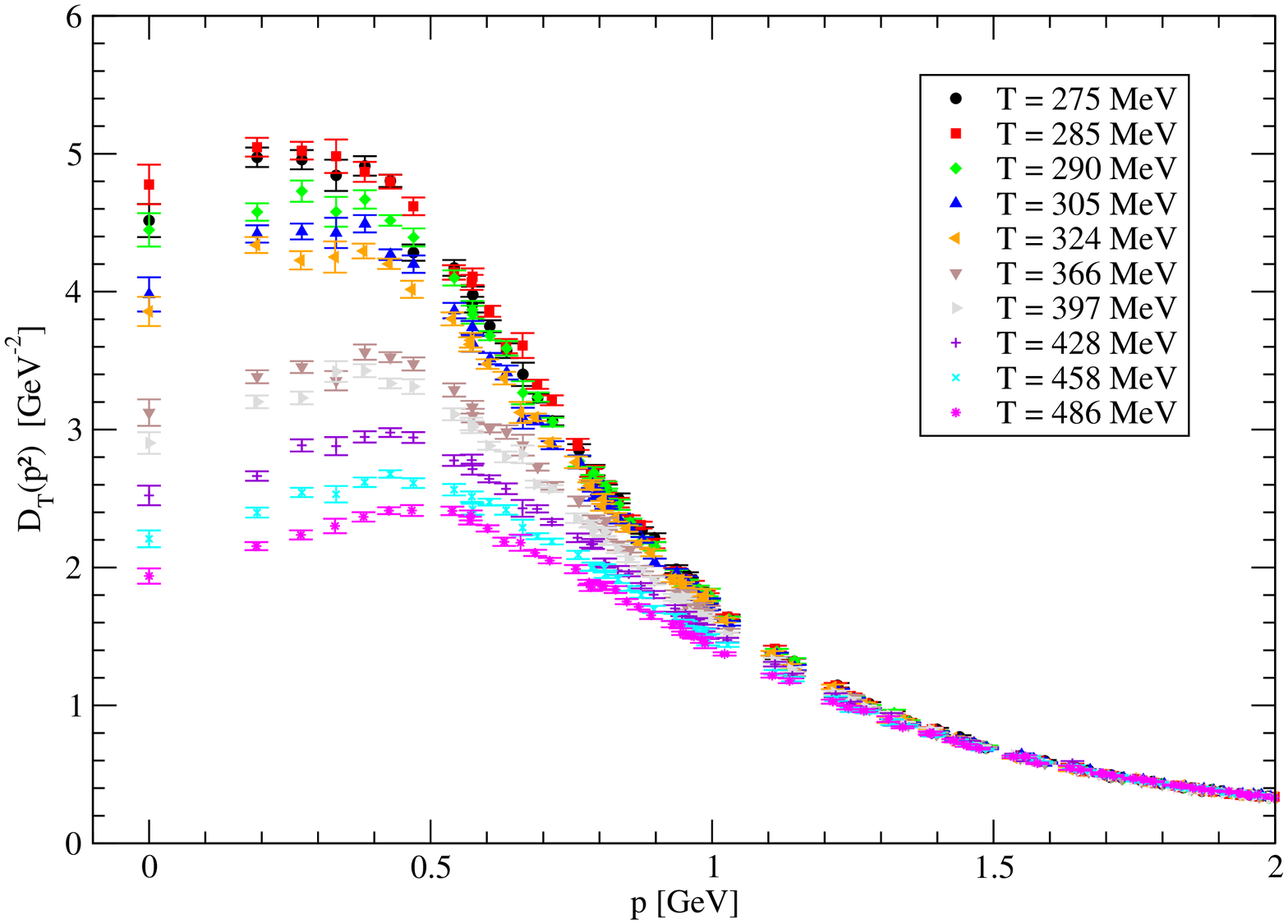} } 
   \end{center}
  \caption{Longitudinal (upper line) and transverse (lower line) gluon propagator form factors as a function of momentum $p$ and temperature $T$
               for a $\sim(6.5\mbox{fm})^3$ spatial lattice volume.}
   \label{fig:3dtemp}
\end{figure*}

%.......................................................................................................................................
%.......................................................................................................................................
%.......................................................................................................................................
\subsection{Gluon propagator definitions at finite $T$}

In the Landau gauge and at finite temperature, the gluon propagator is described by two tensor structures, 
\bea
D^{ab}_{\mu\nu}(p)  & = & \langle A_\mu^a(p) A_\nu^b(-p)\rangle \nonumber \\
                                &  = & \delta^{ab}\Bigg\{ P^{T}_{\mu\nu} \, D_{T}(p_4,\vec{p}) + P^{L}_{\mu\nu} \, D_{L}(p_4,\vec{p}) \Bigg\} 
\label{tens-struct}
\eea
where Latin letters stand for color indices and Greek letters for space-time indices. The transverse and longitudinal projectors are defined by
\bea
P^{T}_{\mu\nu} &=& (1-\delta_{\mu 4})(1-\delta_{\nu 4})\left(\delta_{\mu \nu}-\frac{p_\mu p_\nu}{\vec{p}^{\,\, 2}}\right) \quad , 
\\ \non
P^{L}_{\mu\nu} &=& \left(\delta_{\mu \nu}-\frac{p_\mu p_\nu}{{p}^{2}}\right) - P^{T}_{\mu\nu} \, .
\label{long-proj}
\eea
The transverse and longitudinal form factors, respectively $D_T$ and $D_L$ are given by
\bea
D_T(p) & =  \frac{1}{2V(N_c^2-1)} \Bigg\{ & \langle A_i^a(p) A_i^a(-p) \rangle    \nonumber \\
            &  &  
            \quad - \frac{p_4^2}{\vec{p}^{\, 2}} \langle A_4^a(p) A_4^a(-p)\rangle \Bigg\} 
 \non \\
D_L(p) &=  \frac{1}{V(N_c^2-1)} &\left( 1+\frac{p_4^2}{\vec{p}^{\, 2}} \right) \langle A_4^a(p) A_4^a(-p)\rangle \ ,
\eea
for $p \ne 0$, and
\bea
D_T(0) & =  \frac{1}{3V(N_c^2-1)}  & \langle A_i^a(0) A_i^a(0) \rangle   \\
D_L(0) &=  \frac{1}{V(N_c^2-1)} & \langle A_4^a(0) A_4^a(0)\rangle \ .
\eea
The momentum space gluon field $A^a_\mu ( \tilde{p} )$ reads
\bea
  A_\mu (\tilde{p}) & = & \sum_x \, e^{-i \tilde{p} ( x + \frac{a}{2} \hat{e}_\mu  ) } \, A_\mu ( x + a \hat{e}_\mu / 2 ) \ , \\
  A_\mu ( x + \frac{a}{2} \hat{e}_\mu  ) & = & \frac{1}{2 i g_0} \Big[ U_\mu(x) - U^\dagger_\mu \Big]  \nonumber \\
           & &  \qquad - \frac{1}{6 i g_0} \mbox{Tr} \Big[ U_\mu(x) - U^\dagger_\mu \Big] 
\eea
where
\be
   \tilde{p}_\mu = \frac{ 2 \, \pi \, n_\mu}{a L_\mu} \ , \qquad\qquad n_\mu = 0, 1, \dots , L_\mu - 1
\ee
and $L_\mu$ is the lattice length over direction $\mu$. For the continuum momentum we take the standard definition
\begin{equation}
   q_\mu = \frac{2}{a} \, \sin \left( \frac{\pi}{L_\mu} \, n \right) \, , \hspace{0.2cm} n = 0, \dots, L_\mu - 1 \, .
\end{equation}

%+++++++++++++++++++++++++++++++++++++++++++++++++++++++++++++++++++++++++++
%+++++++++++++++++++++++++++++++++++++++++++++++++++++++++++++++++++++++++++
%+++++++++++++++++++++++++++++++++++++++++++++++++++++++++++++++++++++++++++
\begin{figure*}[th!] %  figure placement: here, top, bottom, or page
\begin{center}
\includegraphics[width=1.0\columnwidth]{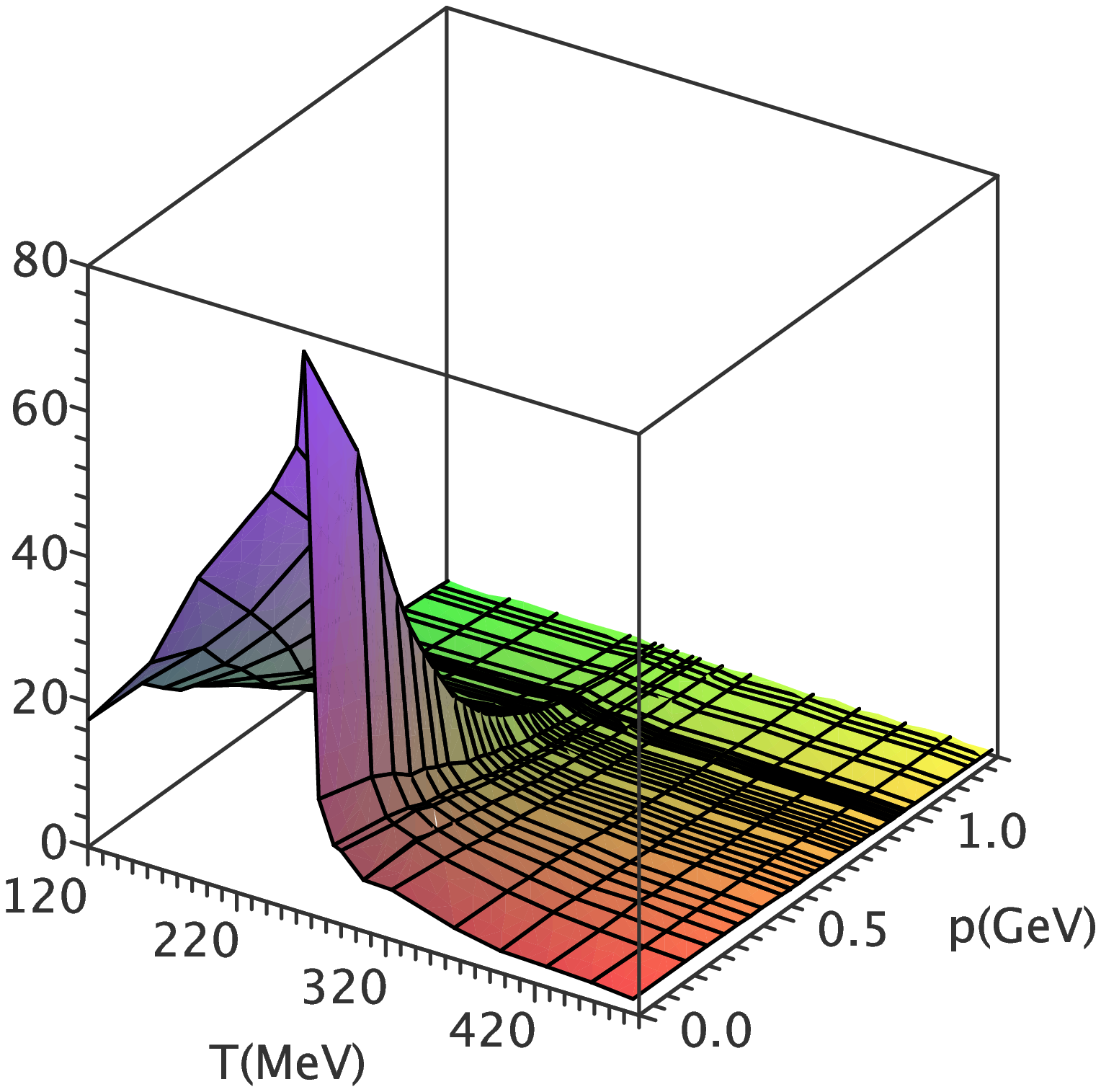}
\includegraphics[width=1.0\columnwidth]{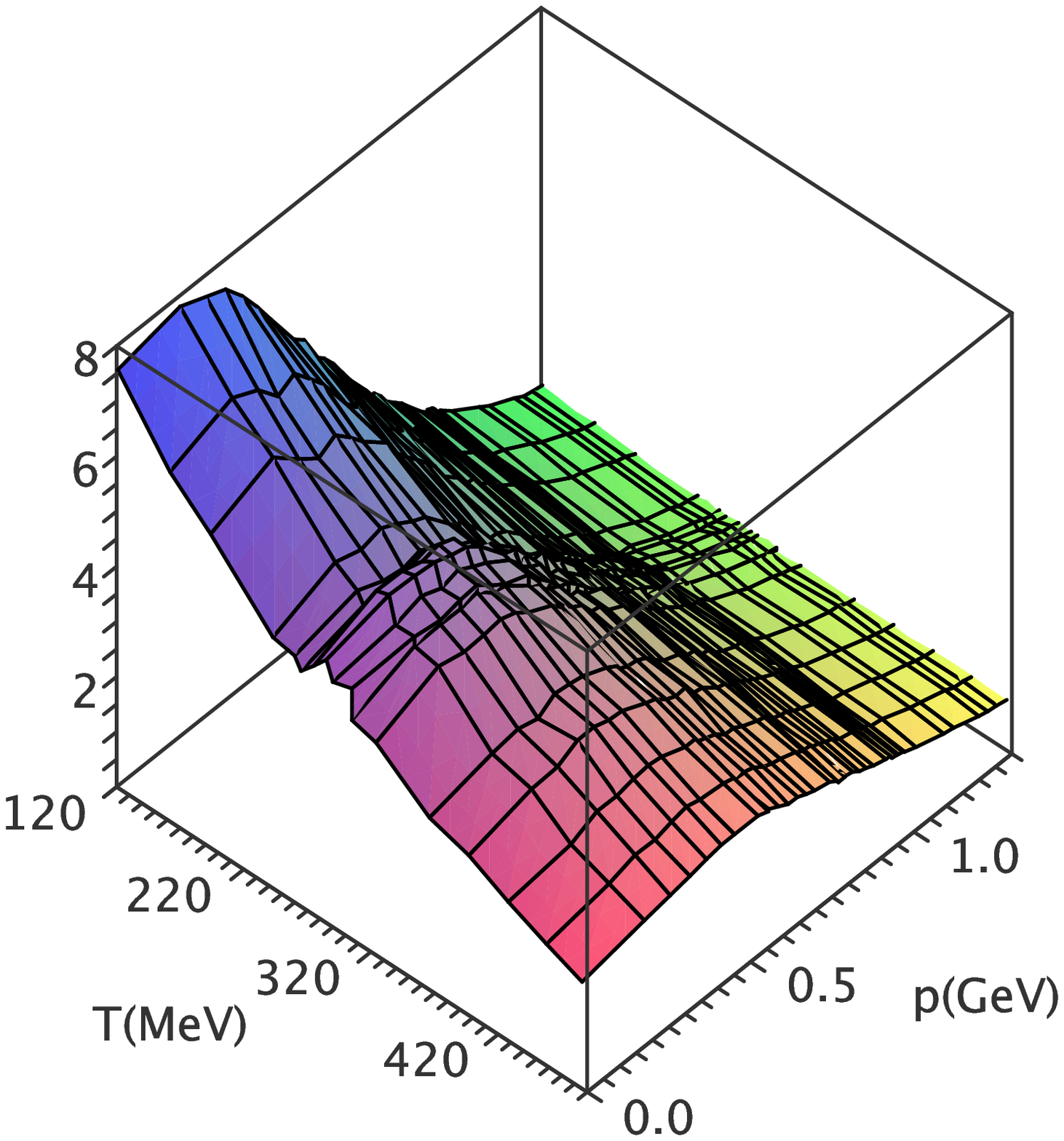}
\end{center}
\caption{Surface plots of the longitudinal (left figure) and transverse (right figure) gluon propagator form factors as a function of momentum $p$
              and temperature $T$.}
\label{fig:translongtemp}
\end{figure*}

%.......................................................................................................................................
%.......................................................................................................................................
%.......................................................................................................................................
\subsection{Renormalization of the Lattice Propagator}

The lattice simulations summarized in Tab.~\ref{tempsetup} were performed using different lattice spacings but the 
same spatial volume $\sim (6.5$ fm$)^3$. In order to compare the data coming from the different simulations, the 
lattice propagator was renormalized as described below.

The bare lattice form factors $D_T(q)$ and $D_L(q)$, after performing the momenta cuts discussed in Section \ref{sistematicos}, are fitted to the one-loop inspired result,
\bea
   D(p^2) = \frac{Z}{p^2} \left( \ln \frac{p^2}{\Lambda^2} \right)^{-\gamma} \, ,
   \label{fit_uv}
\eea  
where $\gamma = 13/22$ is the gluon anomalous dimension, including only the momenta $p \geq p_{min}$, where $p_{min}$ is the
smallest momentum such that the $\chi^2/d.o.f.$ of fit to Eq. (\ref{fit_uv}) is 
smaller than 1.8. 
Then, we use the fit results to compute the renormalization constant $Z_R$ such that
\bea
    D (p^2)  & = & Z_R D_{Lat} (p^2) \, ,
\label{renormalized}
\eea
and requiring the renormalized propagator to verify
\bea
  \left.    D(p^2) \right|_{p^2 = \mu^2} = \frac{1}{\mu^2} \, .
\eea
For the renormalization scale we take $\mu = 4$ GeV $ > p_{min}$. All data shown below refers to renormalized form 
factors.

In what concerns renormalization, $D_L$ and $D_T$ were treated separately. It turned out that, 
for the same temperature, the $Z_R$ associated with $D_L$ and with $D_T$ are compatible within 
one standard deviation.

%+++++++++++++++++++++++++++++++++++++++++++++++++++++++++++++++++++++++++++
%+++++++++++++++++++++++++++++++++++++++++++++++++++++++++++++++++++++++++++
%+++++++++++++++++++++++++++++++++++++++++++++++++++++++++++++++++++++++++++
\begin{figure}[t!] %  figure placement: here, top, bottom, or page
   \vspace{0.3cm}
   \centering
   \includegraphics[width=1.0\columnwidth]{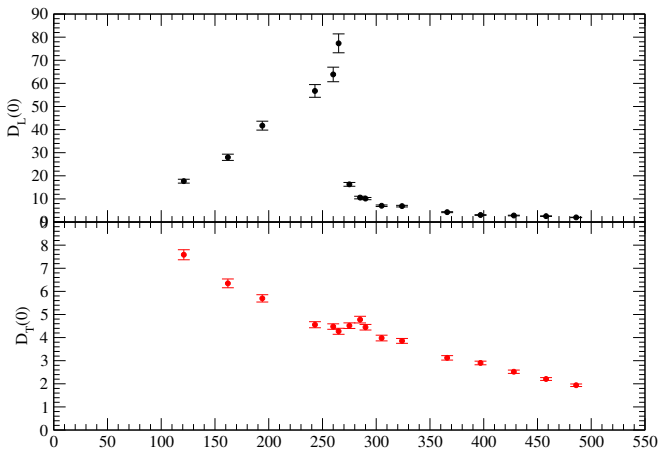}
\caption{Values of the electric and magnetic form factors at zero momentum, as a function of the temperature.}
\label{fig:dzero}
\end{figure}

%+++++++++++++++++++++++++++++++++++++++++++++++++++++++++++++++++++++++++++
%+++++++++++++++++++++++++++++++++++++++++++++++++++++++++++++++++++++++++++
%+++++++++++++++++++++++++++++++++++++++++++++++++++++++++++++++++++++++++++
\begin{figure*}[t!] %  figure placement: here, top, bottom, or page
\begin{center}
\includegraphics[width=0.85\columnwidth]{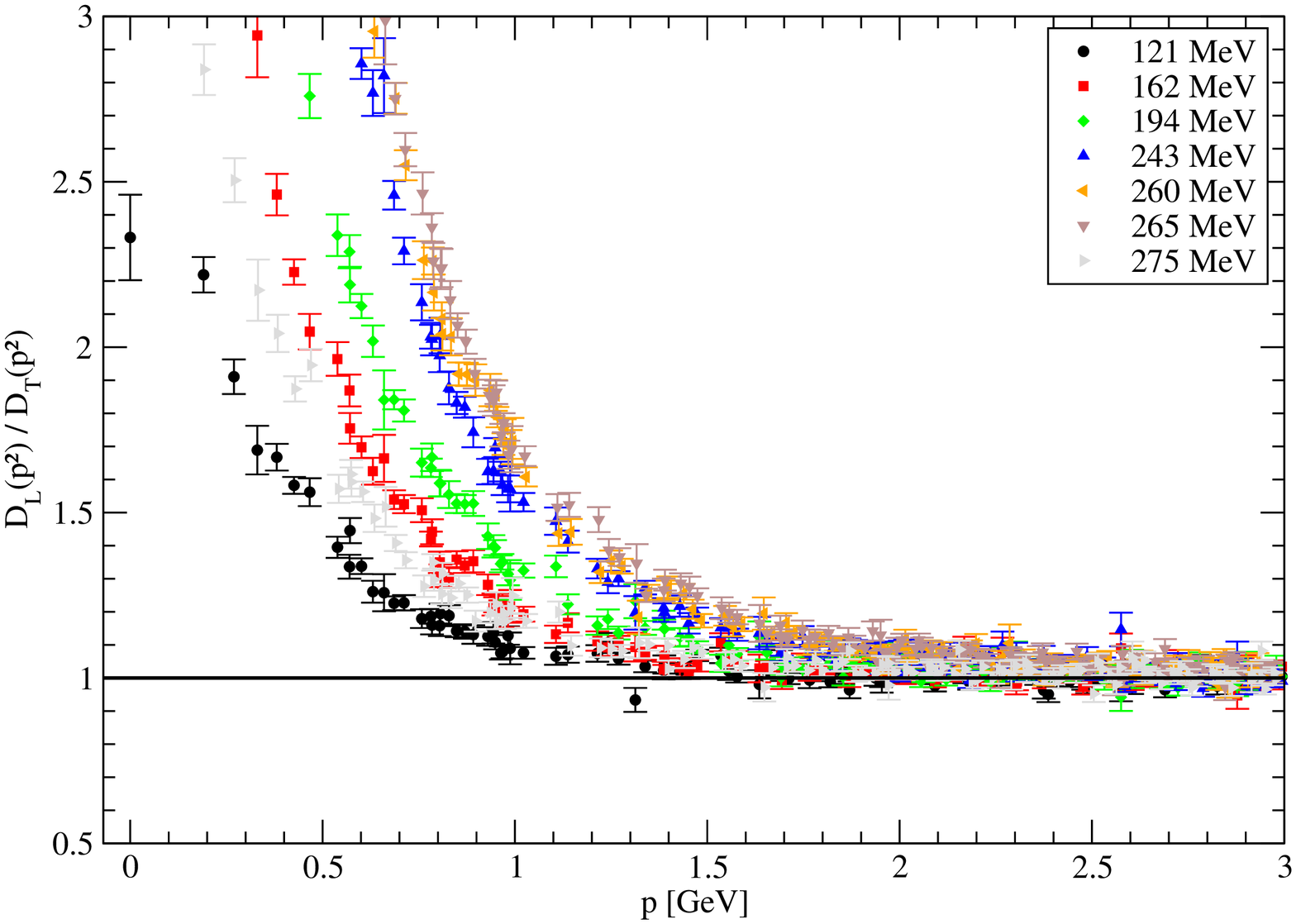} \qquad
\includegraphics[width=0.85\columnwidth]{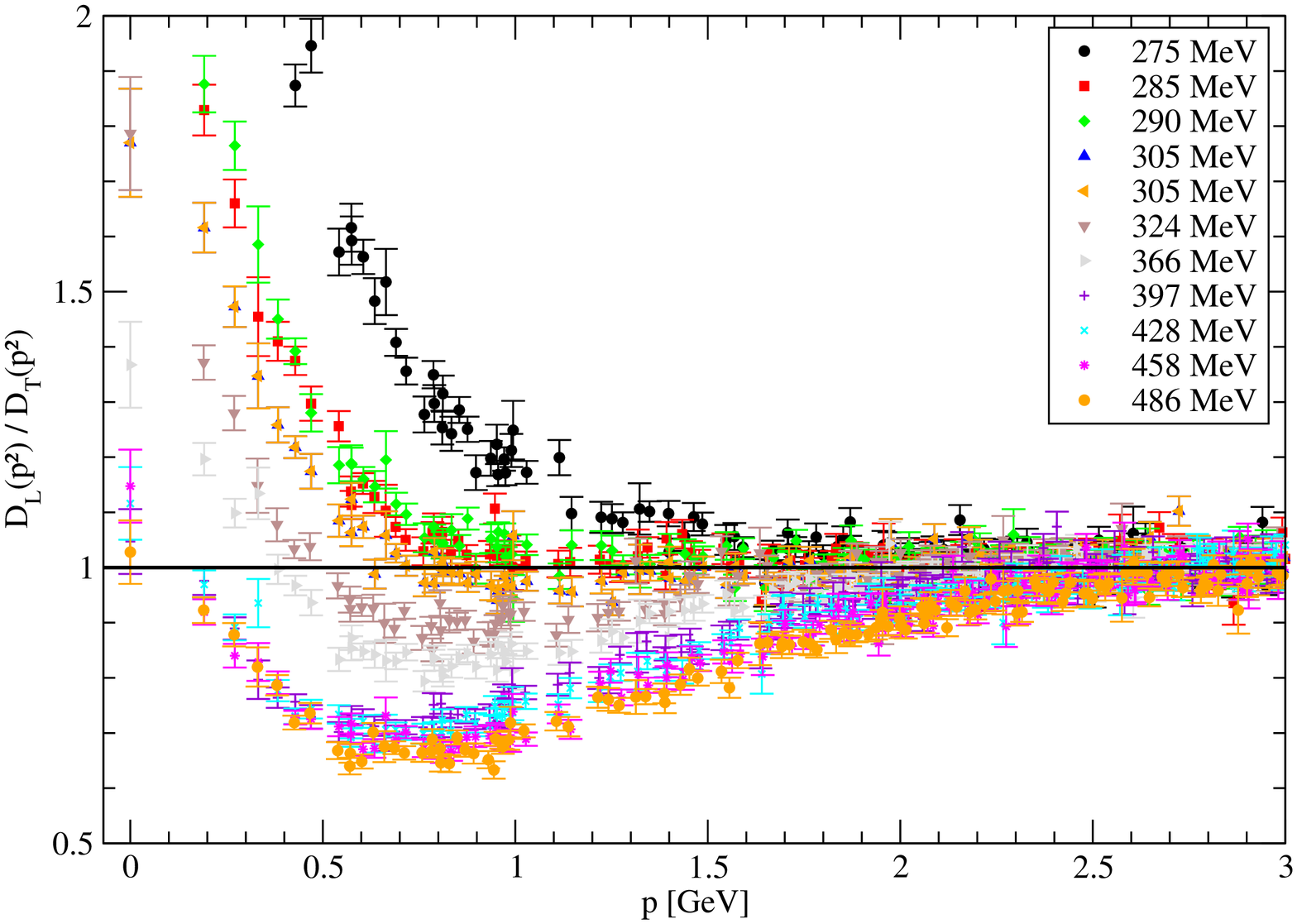}
\end{center}
\caption{$D_L (p^2) / D_T (p^2)$ as a function of the temperature.}
\label{fig:ratios_T}
\end{figure*}

%.......................................................................................................................................
%.......................................................................................................................................
%.......................................................................................................................................
\subsection{Systematic Effects \label{sistematicos}}

The simulations are performed on an hypercubic lattice which breaks rotational invariance. 
In order to reduce lattice spacing effects we have performed the cuts described 
in~\cite{Aouane:2011fv,Leinweber:1998uu} for momenta above 1 GeV. For momenta below 1 GeV, we consider all
the lattice data. In principle, the renormalization procedure with the momentum cuts removes all lattice spacing 
effects from lattice data. The results discussed in~\cite{Oliveira:2012hx,Silva:2013ak} corroborate this. 

In the same set of papers, the longitudinal and transverse form factors show a moderate dependence on the lattice 
volume. Due to limited computational power, our strategy to minimize finite volume effects was to consider a fixed 
physical volume of $\sim (6.5$ fm$)^3$. Given the large physical volume, we expect the finite volume effects to be 
small.  For a discussion on the interplay between the finite volume and finite lattice
spacing effects on the gluon propagator at zero temperature see~\cite{Oliveira:2012eh}.

At finite temperature and above $T_c$ the center symmetry is spontaneously broken. Therefore, we expect a
dependence of the gluon propagator on which $Z(3)$-sector the Polyakov loop belongs to~\cite{Damm:1998pd}. 
To overcome this problem, in the simulations in the deconfined phase (where $P_L \neq 0$) a Z(3) flip with respect to 
the temporal direction was performed, 
such that the phase of the Polyakov loop average is in the interval $]-\pi/3, \pi/3]$.

Another source of systematics are Gribov copies, i.e. configurations which satisfy the Landau gauge condition but are 
related by finite gauge transformations. 
This is a difficult and very demanding computational problem for the lattice practitioner. However, the known SU(3) 
lattice simulations show that Gribov copies 
do not change significantly the gluon propagator for the momenta considered here, 
i.e. that the effect due to the copies are, typically, within the statistical error; 
see, for example, \cite{Silva:2004bv,Sternbeck:2012mf}. Due to the limited computational power, in the present work we will not take into account
possible effects due to Gribov copies.

%.......................................................................................................................................
%.......................................................................................................................................
%.......................................................................................................................................
\section{gluon propagator at finite $T$ \label{resultados_finite_T}}

The form factors $D_L(p^2)$ and $D_T(p^2)$, for $p_4 = 0$, as a function of the momentum $\vec{p}^{\ 2}$ 
and temperature $T$ are reported in Figs.~\ref{fig:3dtemp} and~\ref{fig:translongtemp}. 
The zero momentum electric $D_T(0)$ and magnetic $D_L(0)$ form factors, as a function of
$T$,  can be seen in Fig.~\ref{fig:dzero}.
Fig.~\ref{fig:ratios_T} shows the ratio $D_L(p^2) / D_T(p^2)$ for the various temperatures which have been
simulated -- see Tab.~\ref{tempsetup}. 
Our results are in line with the simulations reported in 
Refs.~\cite{Fischer:2010fx,Cucchieri:2011di,Cucchieri:2012nx,Aouane:2011fv}. 

The two gluon propagator form factors show quite different behaviors with $T$ and $p$. 
The electric form factor $D_L(p^2)$ increases as one approaches the critical temperature from below and is strongly suppressed after $T$ crosses $T_c$. This is clearly seen in the left upper plot of Fig.~\ref{fig:3dtemp}. For temperatures above the critical temperature $D_L(p^2)$ decreases
monotonically as $T$ takes higher values -- see right upper plot in Fig.~\ref{fig:3dtemp}. Despite this drastic suppression of $D_L$ as $T$ increases,
there is no obvious qualitative change of the function $D_L(p^2)$ with $T$ and, for each $T$, the electric propagator is a decreasing function of 
the momentum.

On the other hand, the magnetic form factor $D_T(p^2)$ is clearly a monotonous decreasing function of $T$. 
Moreover, although $D_T$ shows large changes with $T$, around  $T_c \sim 270$ MeV it has the opposite behavior of $D_L$.
Around $T_c$, the lower plots in Fig.~\ref{fig:3dtemp} seem to indicate that the derivative essentially vanishes, increasing to
$\partial D_T (p^2) / \partial T \sim 0$. Further, above $T_c$ the functional form of $D_T(p^2)$ is different than
 for $T < T_c$. Indeed, above the critical temperature the magnetic propagator shows a turnover, with a 
 maximum around $p \sim 500$ MeV, not seen in the lowest temperatures. 

The different behavior of the form factors with $T$ is illustrated in Fig.~\ref{fig:dzero} for zero momentum. The figure shows that
$D_L(0)$ and $D_T(0)$ are sensitive to the confinement-deconfinement transition. Furthermore, it also exhibits the relative order of magnitude of
the electric and magnetic propagators which 
is detailed in Fig.~\ref{fig:ratios_T}.
In the infrared region $D_L(p^2)$ is larger than $D_T(p^2)$ for the range of 
temperatures that we have accessed. However, for temperatures above $\sim$ 360 MeV $D_L$ and $D_T$ have 
similar magnitudes close to zero momentum, with $D_T$ becoming larger as $p$ increases. 
For example, for the largest temperatures considered here, $D_T$ is 
about 30\% larger than $D_L$ around $p \sim 600$ MeV. At higher momenta $D_L$ and $D_T$
become identical for all $T$. 
This behavior for sufficiently high $p$ is not unexpected, as
the perturbative propagator should be recovered in the limit where $p \gg 1$.

%%%%%%%%%%%%%%%%%%%%%%%%%    Fits   %%%%%%%%%%%%%%
%.......................................................................................................................................
%.......................................................................................................................................
%.......................................................................................................................................
\section{Gluon Mass as a function of $T$ \label{calculo_gluon_mass}}

In this section, we consider different definitions for the electric and the magnetic gluon mass as a 
function of the temperature, extending the definitions in Refs. 
~\cite{Heller:1997nqa, Nakamura:2003pu, Maas:2011ez}. 
As discussed below, this will allow us to identify possible order parameters for the 
confinement-deconfinement phase transition.

%.......................................................................................................................................
%.......................................................................................................................................
\subsection{Yukawa Mass}

%+++++++++++++++++++++++++++++++++++++++++++++++++++++++++++++++++++++++++++
%+++++++++++++++++++++++++++++++++++++++++++++++++++++++++++++++++++++++++++
%+++++++++++++++++++++++++++++++++++++++++++++++++++++++++++++++++++++++++++
\begin{figure}[t!] %  figure placement: here, top, bottom, or page
   \vspace{0.3cm}
   \centering
   \includegraphics[width=1.0\columnwidth]{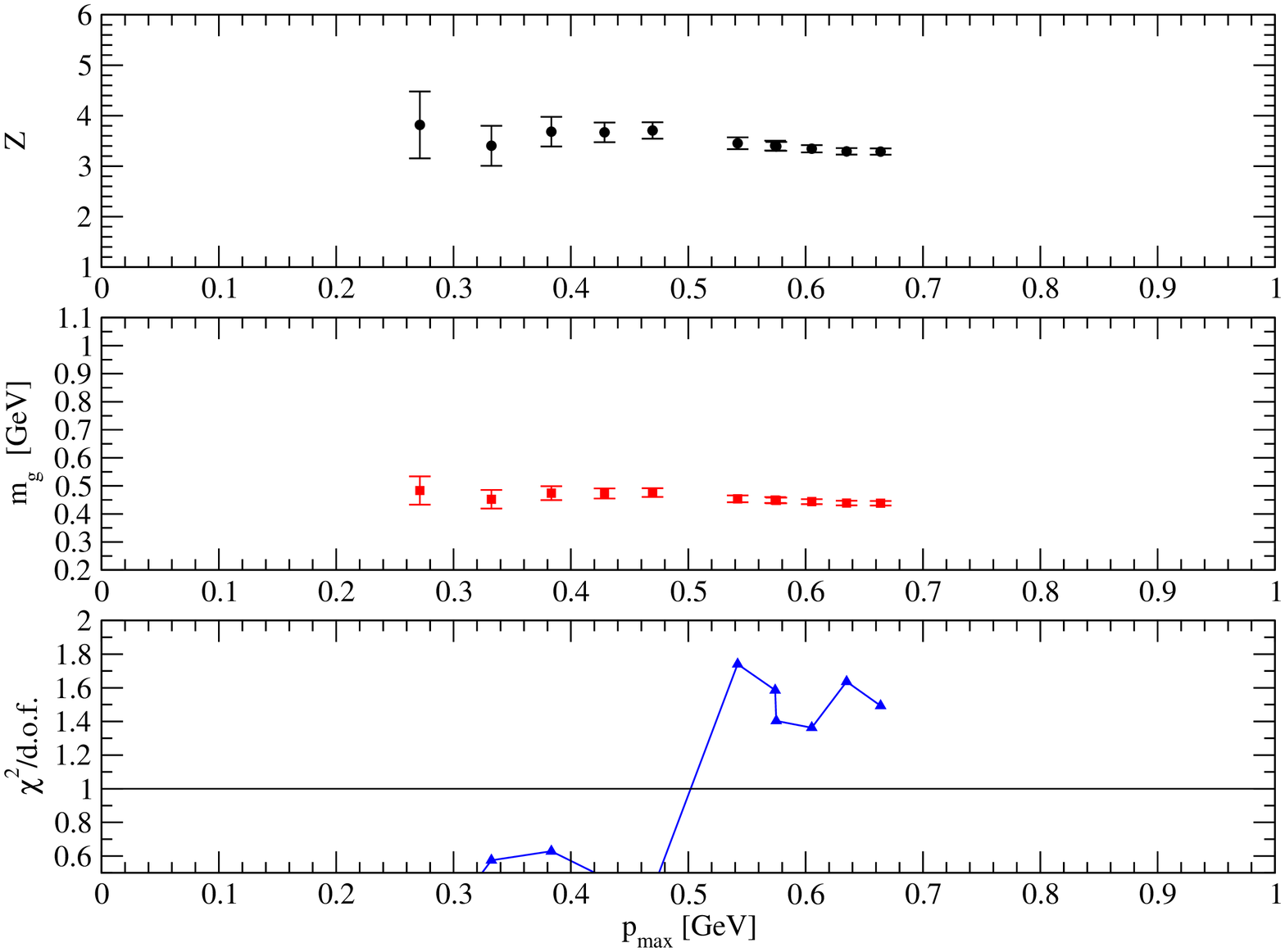}
   \caption{Evolution of the fitting parameters with $p_{max}$ for $T = 275$ MeV.}
\label{fig:yukawa_275}
\end{figure}

In a quasi-particle description of the gluon as used, for example, in constituents models above $T_c$, the gluon is treated as a massive boson. In this
picture, the propagator has a Yukawa form
\be
D(p^2) = {Z \over p^2 + m^2_g} \ ,
 \label{Eq:Yukawa_Z}
\ee
where $m_g$ is the gluon mass and $Z^{1/2}$ the overlap between the gluon state and the quasi-particle massive state.

The values for $Z$ and $m_g$ can be read from fitting the lattice data to (\ref{Eq:Yukawa_Z}) from $p = 0$ up to 
$p_{max}$. 
The upper limit of the fitting interval should be determined in order to a have a $\chi^2/d.o.f. \sim 1$ and to ensure 
that the fitting parameters $Z$ and $m_g$ are independent of the upper bound of the fitting interval. 

In general the gluon mass should be momentum dependent, $m_g = m_g(p^2,T)$, however if $m_g(p^2,T)$ is a slowly changing function of $p^2$, it is possible to identify an interval
of momenta where $m_g(p^2,T)$ is constant and, in this way, define an infrared mass scale. For the gluon propagator at zero 
temperature such an analysis was performed in~\cite{Oliveira:2010xc}, where a $m_g = 648(7)$ MeV, a $Z = 4.044(78)$ for
a $p_{max} \sim 500$ MeV was found.

The magnetic propagator is clearly not described by a Yukawa type propagator for any of temperatures 
considered in this paper. Indeed, fitting $D_T(p^2)$ lattice data to (\ref{Eq:Yukawa_Z}) give always
a $\chi^2/d.o.f.$ which is well above 2 for all the temperatures. 
We conclude that the magnetic propagator does not behave as a quasi-particle massive boson for the 
$T \lesssim 500$ MeV.

%+++++++++++++++++++++++++++++++++++++++++++++++++++++++++++++++++++++++++++
%+++++++++++++++++++++++++++++++++++++++++++++++++++++++++++++++++++++++++++
%+++++++++++++++++++++++++++++++++++++++++++++++++++++++++++++++++++++++++++
\begin{table}[t!]
\begin{center}
\begin{tabular}{c@{\hspace{0.5cm}}l@{\hspace{0.5cm}}l@{\hspace{0.5cm}}l@{\hspace{0.4cm}}l}
\hline
 Temp.           & $p_{max}$ & $Z$            &  $m_g(T)$ & \multicolumn{1}{c}{$\chi^2/d.o.f.$} \\
 (MeV)          &   (GeV)                 &                   &   (GeV)       &  \\
 \hline
 121              & 0.467          & 4.28(16)    & 0.468(13)                                   &   1.91 \\
 162              & 0.570          &  4.252(89) & 0.3695(73)                                 &   1.66 \\
 194              & 0.330          &  5.84(50)   & 0.381(22)                                   &   0.72 \\
 243              &  0.330          &  8.07(67)  & 0.374(21)                                   &   0.27 \\
 260              &  0.271          &  8.73(86)  & 0.371(25)                                   &   0.03 \\
 265              &  0.332          &  7.34(45)  & 0.301(14)                                   &   1.03  \\
 275              &  0.635          &  3.294(65) &  0.4386(83)                               &   1.64  \\
 285              &  0.542          &  3.12(12)   & 0.548(16)                                  &   0.76  \\
 290              &  0.690          &  2.705(50) & 0.5095(85)                                &   1.40  \\
 305              &  0.606          &  2.737(80) & 0.5900(32)                                &   1.30 \\
 324              &  0.870          &  2.168(24) & 0.5656(63)                                &   1.36 \\
 366              &  0.716          &  2.242(55) & 0.708(13)                                  &   1.80 \\
 397              &  0.896          &  2.058(34) & 0.795(11)                                   &  1.03 \\
 428              &  1.112          &  1.927(24)  & 0.8220(89)                                &  1.30 \\
 458              & 0.935          &  1.967(37)  & 0.905(13)                                   & 1.45  \\
 486              & 1.214          &  1.847(24)  & 0.9285(97)                                 & 1.55  \\
\hline
\end{tabular}
\end{center}
\caption{Results of fitting the longitudinal propagator $D_L(p^2)$ to the Yukawa (\ref{Eq:Yukawa_Z}) from $p = 0$ up to $p_{max}$.}
\label{Tab:Yukawa_fits}
\end{table}

On the other hand the electric propagator is well described in the infrared region by a Yukawa propagator. 
In Fig.~\ref{fig:yukawa_275} we show how the fitting parameters and the quality of the fit, measured by the 
$\chi^2/d.o.f.$, changes with $p_{max}$ for $T = 275$ MeV. Similar curves can be shown for the other temperatures.

The outcome of fitting the $D_L(p^2)$ lattice data to Eq. (\ref{Eq:Yukawa_Z}) are reported in 
Tab.~\ref{Tab:Yukawa_fits} and are shown in Fig.~\ref{Fig:z_m_yukawa}. 
The curve in Fig.~\ref{Fig:z_m_yukawa} is the fit of the measured
gluon mass to the prediction of the perturbative~\cite{Arnold:1995bh} functional form of Eq. (\ref{Eq:perturbative}),
extended with a constant term to account for possible non-perturbative corrections,
\be
  m_g (T) = a + b \, T  \, .
  \label{Eq:Pert_Fit}
\ee
If one excludes the data point for $T = 428$ MeV, which seems to be slightly below the other data points for 
$m_g(T)$, Eq. (\ref{Eq:Pert_Fit}) gives a good description of the gluon mass for $T \geq 397$ MeV. 
The measured parameters are $a = 193(88)$ MeV, $b = 1.52(20)$
for a $\chi^2/d.o.f. = 1.50$, where $T$  and $m_g(T)$ are in MeV. 

We would like to call the reader attention for the good agreement between $m_g = 0.32 \pm 0.07$ GeV
estimated from experimental data for heavy ions~\cite{Bicudo:2012wt} and our estimate for $T = 265$ MeV
where $m_g = 0.301 \pm 0.014$ GeV -- see Tab. \ref{Tab:Yukawa_fits}.

We call the reader attention that in some earlier studies of gluon screening masses, 
like in ~\cite{Heller:1997nqa,Nakamura:2003pu}, the prop-
agator is computed in coordinate space, and a position
space Yukawa form $G(z) = C \, exp(-m_g z)$ is assumed. This is equivalent, in our approach, to consider
(\ref{Eq:Yukawa_Z}) as an approximation to the gluon form factors.
The masses are computed via the use of a point-to-all propagator and
this procedure results in higher statistical errors. While the
results of ~\cite{Heller:1997nqa} are for Landau gauge but for the gauge
group SU(2), ~\cite{Nakamura:2003pu} uses mainly the Feynman gauge and thus a direct
comparison with our results should be done with care.

%.......................................................................................................................................
%.......................................................................................................................................
\subsection{Running Gluon Mass}

In~\cite{Oliveira:2010xc} the zero temperature lattice gluon propagator for pure Yang-Mills SU(3) gauge theory was described via a running gluon mass
which also describes the decoupling type of solution of the Dyson-Schwinger
equations~\cite{Cornwall:1981zr,Aguilar:2008xm}. A generalized
functional form for the running mass is able to described the gluon propagator over the full range of momenta 
and all $T$ considered here.

\begin{figure}[t!] %  figure placement: here, top, bottom, or page
   \centering
   \includegraphics[width=1.0\columnwidth]{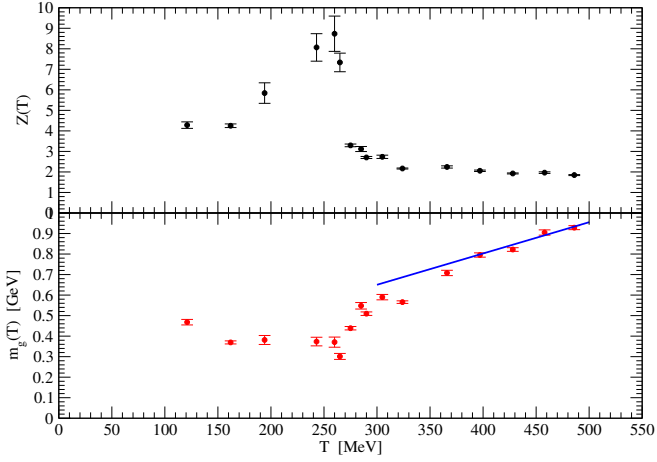}
   \caption{$Z(T)$ and $m_g(T)$ from fitting the longitudinal gluon propagator to Eq. (\ref{Eq:Yukawa_Z}). The curve in the lower figure is
                 the fit of $m_g$ to the functional form predicted by the perturbation theory -- see text for details.}
\label{Fig:z_m_yukawa}
\end{figure}

Let us consider the following functional form for the gluon propagator
\begin{equation}
  D(p^2) = \frac{ Z(p^2; T) }{p^2 + M^2(p^2;T)} ,
  \label{Eq:Fitcomplicado}
\end{equation}  
where
\begin{equation}
  Z(p^2; T )  =  z_0 \left[ \ln \frac{p^2 + r(T) \,  M^2(p^2;T) }{\Lambda^2 (T)} \right]^{-\gamma} \ ,
\end{equation}
$\gamma = 13/22$ is the anomalous gluon dimension and
\begin{equation}  
  M^2(p^2;T)  =  \frac{m^4_0(T)}{p^2 + m^2_1(T)}
\end{equation}
is the temperature dependent running gluon mass. Note that,  if one takes $m_1 = m_0$ and ignores the temperature dependence, the above expression is the same functional form used in~\cite{Oliveira:2010xc}. While in the $T \to 0$ limit it is sufficient ~\cite{Oliveira:2010xc} to consider a single massive parameter $m_0$, for the best fit at finite $T$ we need to consider $m_1$ and $m_0$ as independent parameters.

\begin{figure}[t!] %  figure placement: here, top, bottom, or page
   \centering
   \includegraphics[width=1.0\columnwidth]{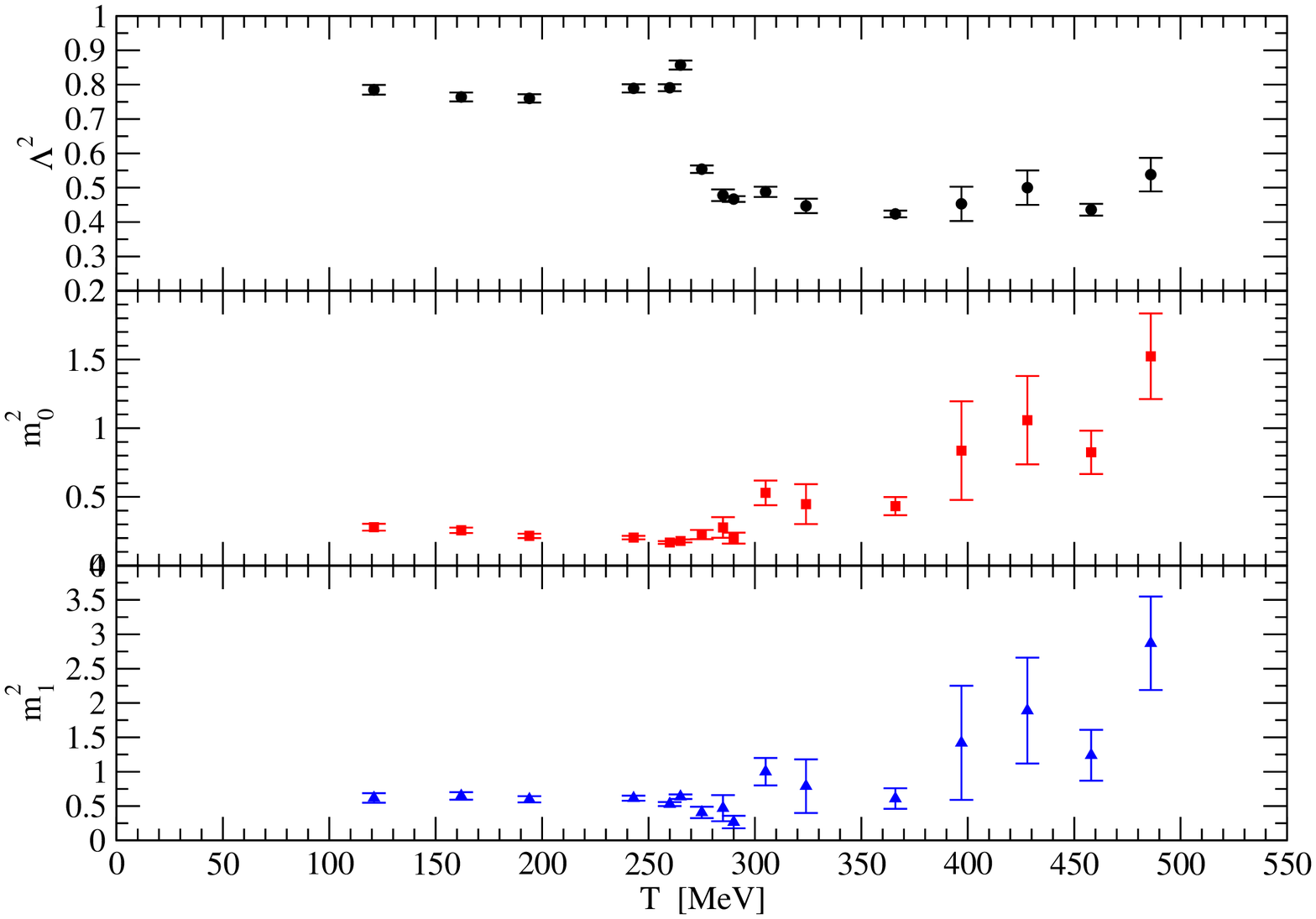}
   \caption{Mass scales, in GeV$^2$, from fitting Eq. (\ref{Eq:Fitcomplicado})  to the lattice $D_L(p^2)$ over the full momentum range.}
\label{fig:massas_complicadas}
\end{figure}

The results of fitting the lattice longitudinal form factor to Eq. (\ref{Eq:Fitcomplicado}) are detailed in Tab. 
\ref{Tab:FitLongFull}. For all
temperatures, $\chi^2/d.o.f.$ is close to unit, i.e. the functional form (\ref{Eq:Fitcomplicado}) is able to describe the 
lattice over the full range of momenta. 
The normalization $z_0 \sim 2$ is essentially independent of $T$. The temperature dependence
of the various mass scales in (\ref{Eq:Fitcomplicado}) is summarized in Fig.~\ref{fig:massas_complicadas}. 
Of the mass scales $\Lambda^2$, $m^2_0$ and $m^2_1$ clearly $\Lambda^2$ is the most sensible to the confinement-deconfinement transition, taking the
value $\Lambda \sim 0.89$ GeV for $T \lesssim T_c$ and $\Lambda \sim 0.67$ GeV for $T \gtrsim T_c$. In what concerns $m^2_0$ and $m^2_1$,
the two mass scales are roughly constant up to  $T \sim 300$ MeV, i.e. slightly above $T_c$, and increase with $T$ for higher temperatures.

%+++++++++++++++++++++++++++++++++++++++++++++++++++++++++++++++++++++++++++
%+++++++++++++++++++++++++++++++++++++++++++++++++++++++++++++++++++++++++++
%+++++++++++++++++++++++++++++++++++++++++++++++++++++++++++++++++++++++++++
\begin{table*}[t!]
\begin{center}
\begin{tabular}{l@{\hspace{0.6cm}}l@{\hspace{0.4cm}}r@{\hspace{0.4cm}}l@{\hspace{0.5cm}}l@{\hspace{0.4cm}}l@{\hspace{0.6cm}}l}
\hline
T   & $z_0$        & \multicolumn{1}{l}{$r$}              &  $\Lambda^2$  &  $m^4_0$    & $m^2_1$   & $\chi^2/d.o.f$    \\
\hline
 121 & 1.9271(67)   & $11.5 \pm 1.5$   &  0.785(14)    &  0.078(14)  & 0.618(69) & 1.28              \\
 162 & 1.9367(61)   & $11.7 \pm 1.3$   &  0.764(13)    &  0.066(10)  & 0.647(54) & 1.18              \\
 194 & 1.9379(59)   & $14.2 \pm 1.4$   &  0.760(12)    &  0.0468(68) & 0.599(45) & 1.26              \\
 243 & 1.9237(57)   & $15.0 \pm 1.4$   &  0.789(12)    &  0.0415(54) & 0.615(37) & 1.47              \\
 260 & 1.9238(50)   & $19.7 \pm 1.6$   &  0.791(10)    &  0.0283(33) & 0.530(29) & 1.43              \\
 265 & 1.8847(65)   & $21.2 \pm 1.8$   &  0.857(13)    &  0.0321(39) & 0.637(34) & 1.45              \\
 275 & 2.0571(66)   & $11.9 \pm 2.9$   &  0.554(11)    &  0.051(15)  & 0.407(83) & 1.12              \\
 285 & 2.105(12)    & $11.5 \pm 5.2$   &  0.478(17)    &  0.077(42)  & 0.47(19)  & 1.10              \\
 290 & 2.1224(59)   & $20.5 \pm 7.7$   &  0.4670(84)   &  0.040(16)  & 0.267(91) & 1.06              \\
 305 & 2.1143(92)   & $4.6 \pm 1.6$   &  0.488(15)    &  0.280(95)  & 1.00(20)  & 1.10              \\
 324 & 2.141(15)    & $8.5 \pm 5.3$   &  0.447(21)    &  0.20(13)   & 0.79(39)  & 1.10              \\
 366 & 2.1684(73)   & $13.8 \pm 4.5$   &  0.4235(97)   &  0.187(57)  & 0.61(15)  & 0.96              \\
 397 & 2.138(34)    & $5.0 \pm 4.8$   &  0.453(50)    &  0.70(60)   & 1.42(83)  & 1.07              \\
 428 & 2.115(30)    & $4.2 \pm 2.9$   &  0.500(50)    &  1.12(68)   & 1.89(77)  & 1.00              \\
 458 & 2.168(13)    & $9.5 \pm 4.3$   &  0.436(17)    &  0.68(26)   & 1.24(37)  & 1.09              \\
 486 & 2.113(27)    & $2.9 \pm 1.6$   &  0.538(49)    &  2.32(95)   & 2.87(68)  & 1.03              \\
\hline
\end{tabular}
\end{center}
\caption{Results of fitting the longitudinal propagator $D_L(p^2)$ to Eq. (\ref{Eq:Fitcomplicado}) for the range of momenta. The temperature is in
MeV, while the mass scales are in powers of GeV.}
\label{Tab:FitLongFull}
\end{table*}

%+++++++++++++++++++++++++++++++++++++++++++++++++++++++++++++++++++++++++++
%+++++++++++++++++++++++++++++++++++++++++++++++++++++++++++++++++++++++++++
%+++++++++++++++++++++++++++++++++++++++++++++++++++++++++++++++++++++++++++

%+++++++++++++++++++++++++++++++++++++++++++++++++++++++++++++++++++++++++++
%+++++++++++++++++++++++++++++++++++++++++++++++++++++++++++++++++++++++++++
%+++++++++++++++++++++++++++++++++++++++++++++++++++++++++++++++++++++++++++
\begin{figure}[t!] %  figure placement: here, top, bottom, or page
   \vspace{0.5cm}
   \centering
   \includegraphics[width=1.0\columnwidth]{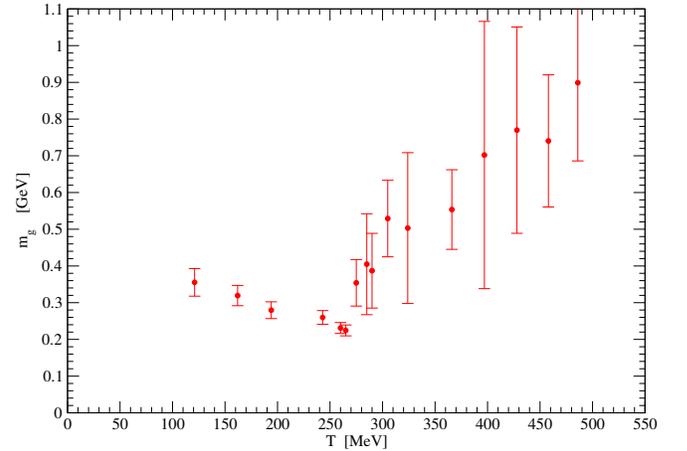}
   \caption{Infrared mass scale associated with $D_L(p^2)$ taken as $m^2_0/m_1$.}
\label{fig:massIR_complicadas}
\end{figure}

The functional form (\ref{Eq:Fitcomplicado}) allows for the definition of an 
infrared mass scale $m^{IR}_g = m^2_0 / m_1$; note that
for low momenta $p^2 \ll  m^2_1$, (\ref{Eq:Fitcomplicado}) reduces to an Yukawa shape with a mass
given by $m^{IR}$. 
$m^{IR}_g(T)$ can be seen in Fig.~\ref{fig:massIR_complicadas} and reproduces the same 
qualitative behavior as observed in the lower part of Fig.~\ref{Fig:z_m_yukawa}. 
Furthermore, we have checked that the temperature dependence of the data on 
Fig.~\ref{fig:massIR_complicadas} for $T > T_c$ is compatible with the perturbative inspired 
behavior resummed in Eq. (\ref{Eq:Pert_Fit}). 
However, given the large statistical errors of the data, the estimated statistical errors on
$a$ and $b$ are quite large (they are above 50\%).

%+++++++++++++++++++++++++++++++++++++++++++++++++++++++++++++++++++++++++++
%+++++++++++++++++++++++++++++++++++++++++++++++++++++++++++++++++++++++++++
%+++++++++++++++++++++++++++++++++++++++++++++++++++++++++++++++++++++++++++
\begin{table*}[t!]
\begin{center}
\begin{tabular}{l@{\hspace{0.6cm}}l@{\hspace{0.4cm}}r@{\hspace{0.4cm}}l@{\hspace{0.5cm}}l@{\hspace{0.4cm}}l@{\hspace{0.6cm}}l}
\hline
T   & $z_0$        & \multicolumn{1}{l}{$r$}              &  $\Lambda^2$  &  $m^4_0$    & $m^2_1$   & $\chi^2/d.o.f$    \\
\hline
 121 & 1.9723(41)   & $26.8 \pm 3.8$   &  0.6922(73)   &  0.0261(39) & 0.143(20) & 1.72              \\
 162 & 2.0067(49)   & $20.8 \pm 3.1$   &  0.6313(79)   &  0.0356(54) & 0.164(23) & 1.89             \\
 194 & 2.0468(44)   & $14.7 \pm 2.1$   &  0.5671(67)   &  0.0535(73) & 0.215(25) & 1.34   \\
 243 & 2.1211(52)   & $11.1 \pm 1.7$   &  0.4614(66)   &  0.0724(97) & 0.236(27) & 1.41             \\
 260 & 2.1547(54)   & $9.8 \pm 1.7$    &  0.4210(66)   &  0.085(12)  & 0.262(32) & 1.56     \\
 265 & 2.1567(66)   & $11.0 \pm 1.6$   &  0.4101(72)   &  0.0734(87) & 0.228(24) & 1.31    \\
 275 & 2.1408(46)   & $10.2 \pm 1.5$   &  0.4379(59)   &  0.0718(92) & 0.230(25) & 1.36    \\
 285 & 2.1359(56)   & $9.1 \pm 1.6$    &  0.4375(66)   &  0.077(12)  & 0.247(31) & 1.29   \\
 290 & 2.1597(41)   & $7.6 \pm 1.3$    &  0.4194(51)   &  0.090(12)  & 0.257(29) & 1.17  \\
 305 & 2.1553(41)   & $8.8 \pm 1.3$    &  0.4241(54)   &  0.088(11)  & 0.249(26) & 1.26  \\
 324 & 2.1617(52)   & $6.6 \pm 1.3$    &  0.4113(60)   &  0.110(17)  & 0.286(34) & 1.32  \\
 366 & 2.1505(50)   & $7.0 \pm 1.2$    &  0.4308(64)   &  0.134(17)  & 0.293(31) & 1.42  \\
 397 & 2.1110(67)   & $7.6 \pm 1.2$    &  0.4693(83)   &  0.142(17)  & 0.302(29) & 1.36  \\
 428 & 2.0921(65)   & $8.6 \pm 1.1$    &  0.5014(89)   &  0.156(15)  & 0.299(24) & 1.45  \\
 458 & 2.0860(61)   & $7.2 \pm 1.0$    &  0.5150(86)   &  0.205(21)  & 0.345(29) & 1.41  \\
 496 & 2.0629(60)   & $8.7 \pm 1.0$    &  0.5543(93)   &  0.210(18)  & 0.330(24) & 1.54 \\
\hline
\end{tabular}
\end{center}
\caption{Results of fitting the transverse propagator $D_T(p^2)$ to Eq. (\ref{Eq:Fitcomplicado}) for the range of momenta. The temperature is in
MeV, while the mass scales are in powers of GeV.}
\label{Tab:FitTransFull}
\end{table*}

The functional form given in Eq. (\ref{Eq:Fitcomplicado}) is able to describe also the magnetic form factor $D_T(p^2)$. In Tab. \ref{Tab:FitTransFull}
we provide the output of the fits of the full range transverse lattice propagator data for the various temperatures. Again, the overall normalization
factor $z_0 \sim 2$ seems to be independent of $T$. $\Lambda$ shows a non-monotonic behavior with $T$, while $m^2_0$ and $m^2_1$ are 
increasing functions of $T$. However, the mass scales $\Lambda^2(T)$, $m^2_0(T)$ and $m^2_1(T)$ do not show a clear indication of the
confinement-deconfinement phase transition.

%+++++++++++++++++++++++++++++++++++++++++++++++++++++++++++++++++++++++++++
%+++++++++++++++++++++++++++++++++++++++++++++++++++++++++++++++++++++++++++
%+++++++++++++++++++++++++++++++++++++++++++++++++++++++++++++++++++++++++++
\begin{figure}[t!] %  figure placement: here, top, bottom, or page
   \centering
   \includegraphics[width=1.0\columnwidth]{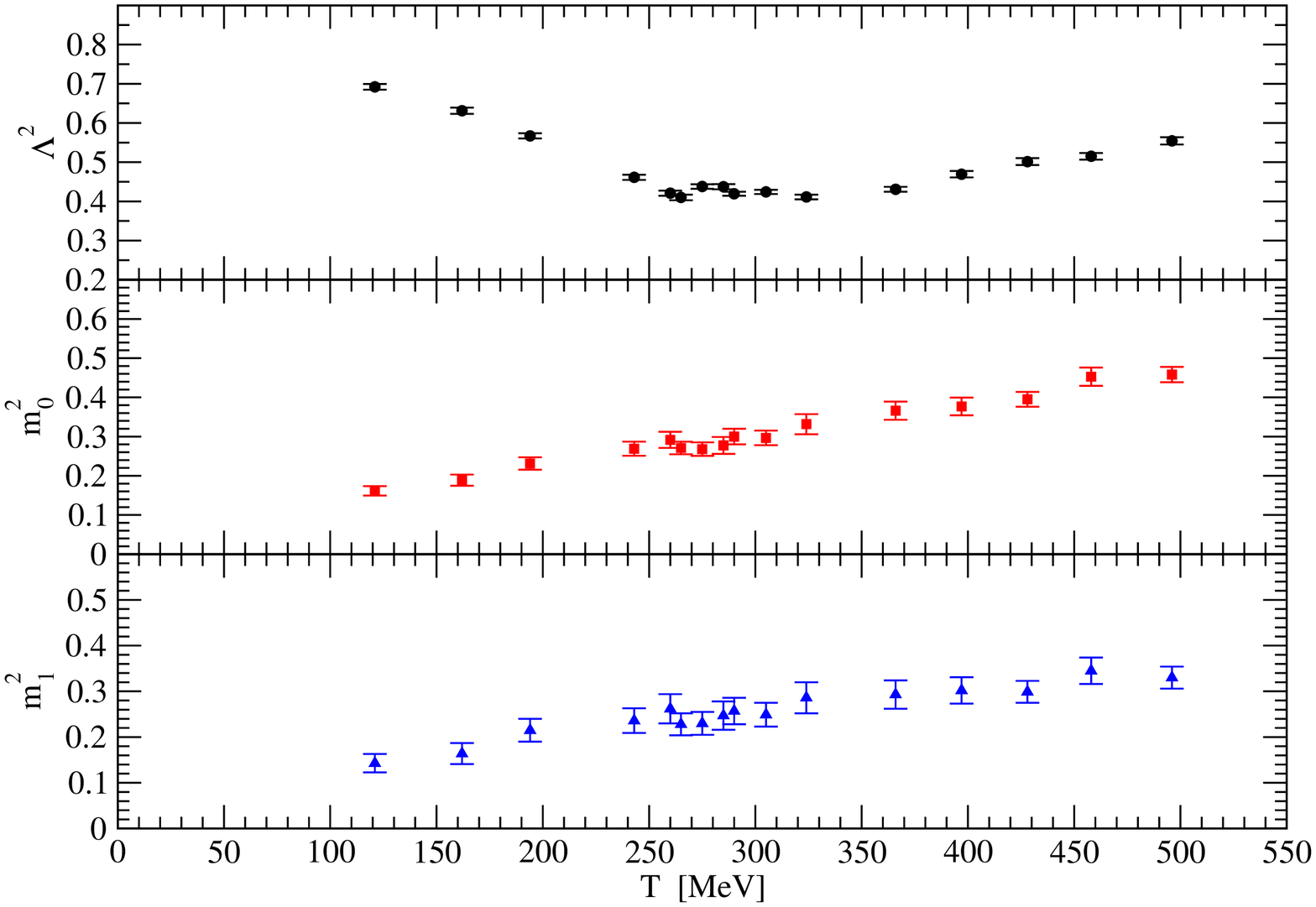}
   \caption{Mass scales, in GeV$^2$, from fitting Eq. (\ref{Eq:Fitcomplicado})  to the lattice $D_T(p^2)$ over the full momentum range.}
\label{fig:massas_Trans_complicadas}
\end{figure}

%+++++++++++++++++++++++++++++++++++++++++++++++++++++++++++++++++++++++++++
%+++++++++++++++++++++++++++++++++++++++++++++++++++++++++++++++++++++++++++
%+++++++++++++++++++++++++++++++++++++++++++++++++++++++++++++++++++++++++++
\begin{figure}[t!] %  figure placement: here, top, bottom, or page
   \vspace{0.5cm}
   \centering
   \includegraphics[width=1.0\columnwidth]{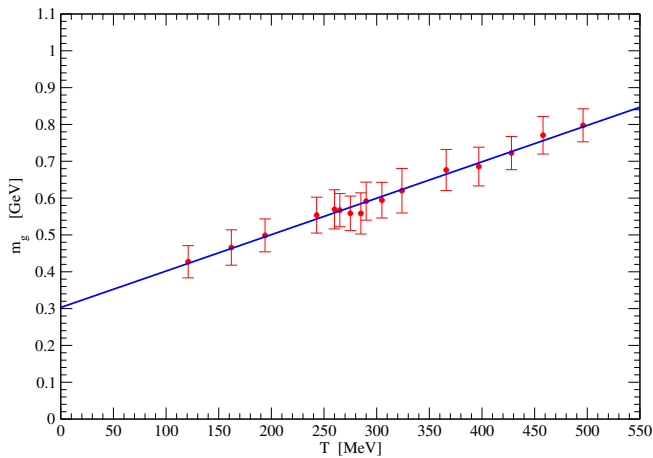}
   \caption{Infrared mass scale associated with $D_T(p^2)$ taken as $m^2_0/m_1$. The full line is the outcome of the linear fit described in the
                 text - see Eq. (\ref{Eq:mIR_trans_linear_fit}).}
\label{fig:massIR_Trans_complicadas}
\end{figure}

Similarly to the electric form factor, one can define an infrared mass scale associated with the magnetic form factor as $m^{IR}_g = m^2_0 / m_1$.
This mass scale, as a function of the temperature, is shown in Fig. \ref{fig:massIR_Trans_complicadas}. The figure points to 
$m^{IR}_g (T)$ as a linear function of $T$. Indeed, a linear fit (full line seen in 
Fig.~\ref{fig:massIR_Trans_complicadas}) to the $m^{IR}_g (T)$ gives
\begin{equation}
 m^{IR}_g (T) =    \big( 303.0 \pm 8.1 \big) + \big( 0.989 \pm 0.025 \big) T
 \label{Eq:mIR_trans_linear_fit}
\end{equation}
for a $\chi^2/d.o.f. = 0.05$ with all dimensionful quantities given in MeV.

A linear behavior of the magnetic screening mass $m_{mag}$ has been used in the perturbative approach to hot
QCD; see, e.g., \cite{Rebhan:1993uz,Alexanian:1995rp,Andersen:2004fp} and references therein.
Furthermore, as already stated in the introduction, the perturbative approach to QCD assumes that the Debye
screening mass associated with the electric form factor is such that, in leading order,
$m_D \sim g T$, while the magnetic mass needed to regulate the perturbative expansion goes as
$m_{mag} \sim g^2 T$. 
In this sense, our results validate, within the statistical accuracy of the simulations, the
functional dependence of the $m_D$ and $m_{mag}$ as taken in perturbation theory.

%.......................................................................................................................................
%.......................................................................................................................................
\subsection{Zero Momentum Mass Scale}

The infrared mass scale $m_D(T)$ associated with the electric form factor $D_L(p^2;T)$ show quite a different 
behavior below and above $T_c$, see Figs.~\ref{Fig:z_m_yukawa} and~\ref{fig:massIR_complicadas}.
Indeed, 
$m_D(T)$ can be used as an order parameter for the confinement-deconfinement phase transition. 
On the other hand, the infrared mass scale $m_{mag}(T)$, linked with the magnetic form factor
$D_T(p^2;T)$, shows a monotonous behavior with $T$, see Fig.~\ref{fig:massIR_Trans_complicadas}, and it is not 
obvious that it can be used to identify the phase transition. 

The definitions of $m_D$ or $m_{mag}$ considered in the previous section requires fitting the lattice data to a 
given functional form either (\ref{Eq:Yukawa_Z}) or (\ref{Eq:Fitcomplicado}). Alternatively, one can use directly the 
lattice propagator data to define a nonperturbative mass scale. A possible choice is to connect the mass scale with 
the deep infrared propagator,
as in Refs. ~\cite{Heller:1997nqa,Maas:2011ez}. 
So let us consider
\be
m(T) = 1 / \sqrt{D( p^2 = 0 ; T)} \ ,
\label{Eq:simpler mass}
\ee
as used in \cite{Maas:2011ez}. 
This choice is equivalent to a Yukawa fit of E.q. (\ref{Eq:Yukawa_Z}) setting the parameter $Z=1$, 
as used e.g. for the high-$T$ region~\cite{Heller:1997nqa} .

\begin{figure}[t]
\begin{center}
    \includegraphics[width=0.95\columnwidth]{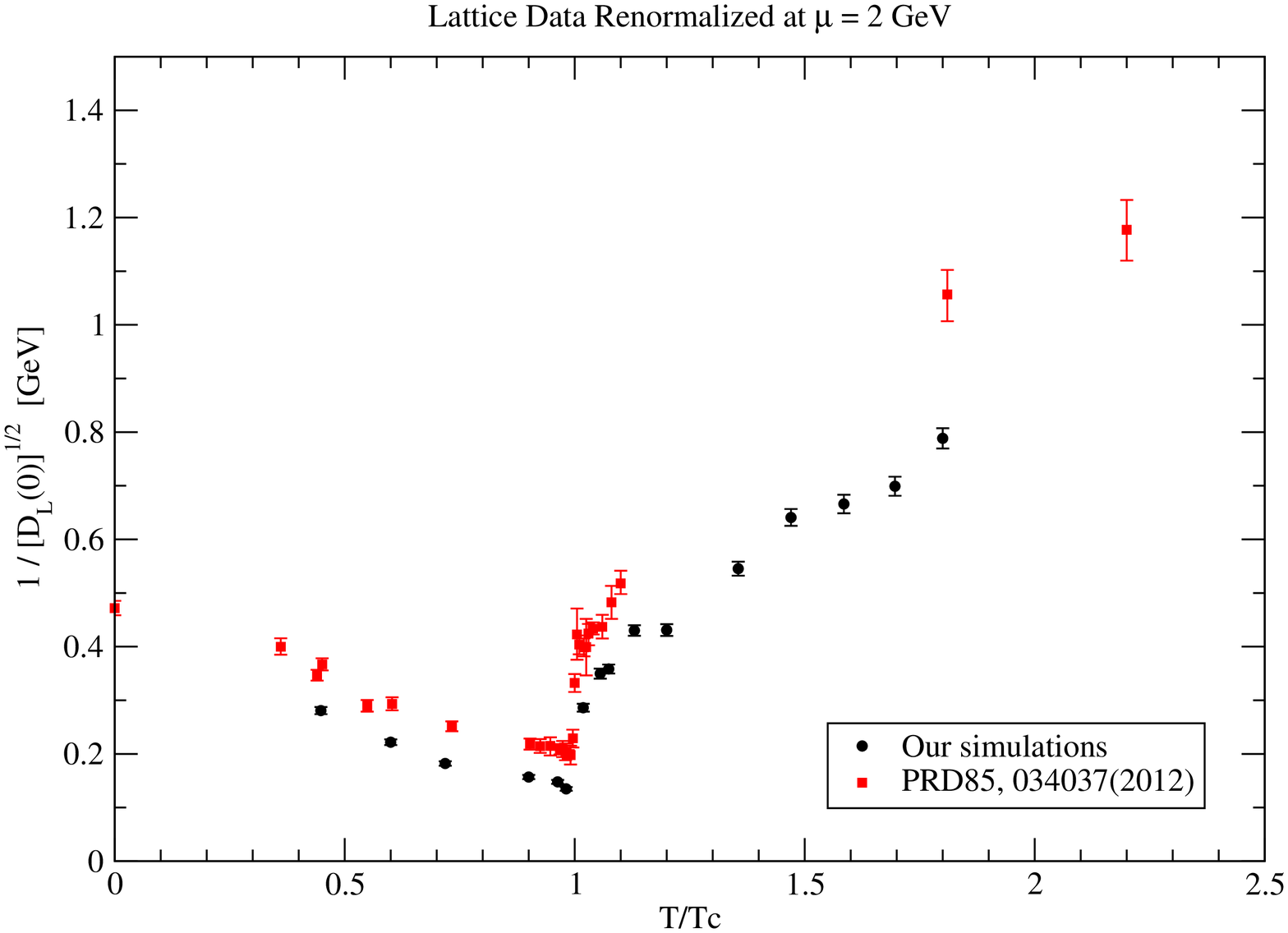}
    \caption{Electric mass defined by Eq. (\ref{Eq:simpler mass}) for the simulations reported in Tab. \ref{tempsetup} renormalized at $\mu = 2$ GeV compared
                  with the data of~\cite{Maas:2011ez}.}%For the data reported in the present work, the x-axis was built taking $T_c = 270$ MeV. The simulations
%                  considered~\cite{Maas:2011ez} take $T_c = 277$ MeV.}
    \label{fig:screeningmass_compare}
\end{center}
\end{figure}

The comparison of our data with~\cite{Maas:2011ez} requires choosing a different renormalization scale, namely $\mu = 2$ GeV. The two sets of data
for the electric form factor $D_L(p^2;T)$, i.e. taking $m = 1 / \sqrt{D_L(0)}$, are compared in Fig. \ref{fig:screeningmass_compare} 
and show essentially the same type of dependence with $T / T_c$. 

The differences between the two sets of data in Fig. \ref{fig:screeningmass_compare} are possibly
due to the finite volume effects and the lattice spacing effects. While the simulations reported here are 
performed considering a spatial volume $V_s \approx (6.5$ fm$)^3$ and lattice spacings smaller than $0.12$ fm, 
those in~\cite{Maas:2011ez} used spatial volumes between $(2.92 - 7.49$ fm$)^3$ and lattice spacings larger 
than $0.16$ fm. While our smallest $\beta$ value is 5.8876, \cite{Maas:2011ez} used $\beta \in [ 5.642 , 5.738 ]$. 
Furthermore, to build Fig.  \ref{fig:screeningmass_compare} we take $T_c = 270$ MeV when using our data,
while~\cite{Maas:2011ez} uses $T_c = 277$ MeV. Given that we do not cover the transition region with 
great detail, this difference on the transition temperature does not change drastically the outcome of 
Fig.  \ref{fig:screeningmass_compare}.

\begin{figure}[t]
\begin{center}
    \includegraphics[width=0.95\columnwidth]{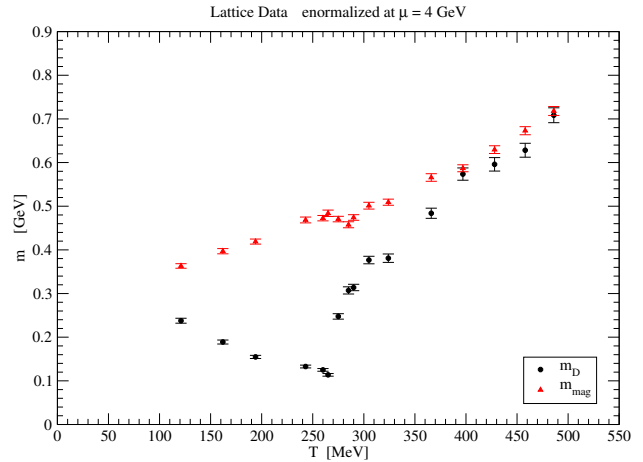}
    \caption{Electric and magnetic mass defined by Eq. (\ref{Eq:simpler mass}) for the simulations reported in Tab. \ref{tempsetup}.}
    \label{fig:screeningmass}
\end{center}
\end{figure}

The electric and magnetic masses defined with Eq. (\ref{Eq:simpler mass}) are compared in 
Fig.~\ref{fig:screeningmass} and reproduce the behavior observed in Figs. ~\ref{Fig:z_m_yukawa}, 
~\ref{fig:massIR_complicadas} and ~\ref{fig:massIR_Trans_complicadas}.
In Ref.  ~\cite{Heller:1997nqa} the large $T$ limit was studied. The authors have shown that the magnetic mass is 
eventually suppressed,  $m_D / m_{mag} \sim \log T$. Here we find the magnetic mass $m_{mag}$ is larger
that the Debye mass when $T < 2 T_c$ . Only at $T \sim 2 T_c$ the electric mass $m_D$ takes over the magnetic 
mass and, therefore, it is only for $T \gtrsim 2 T_c$ that the transverse degrees of freedom become dominant. 
Thus, for the temperatures considered here, $T \in [0, 2 T_c]$, the low temperature degrees of freedom associated
with the gluon longitudinal degrees of freedom are still relevant, and the perturbative magnetic gluon gas
 does not completely set in yet.

Note that at  $T =T_c = 270$ MeV, the electric mass $m_D$ reaches its minimum value. 
Given that we do not cover the transition region with great detail, it is not clear from our data whether  
the mass is discontinuous or its derivative is discontinuous. Nevertheless, when $T$ crosses $T_c$, 
our data for the electric mass  $m_D$ is consistent with a phase transition.

%%%%%%%%%%%%%%%%%%%%%%%%%    Conclusions   %%%%%%%%%%%%%%%%%%%%%%%%%%%%%
%.......................................................................................................................................
%.......................................................................................................................................
%.......................................................................................................................................
%.......................................................................................................................................
\section{Summary and Conclusions \label{conclusoes}}

We have computed the gluon propagator at finite temperature using lattice QCD simulations for pure gauge 
SU(3) Yang-Mills theory on a large spatial volume $V_s \approx ( 6.5$ fm$)^3$. 
Our results are similar to those found by different 
authors~\cite{Fischer:2010fx,Cucchieri:2011di,Cucchieri:2012nx,Aouane:2011fv}.

%+++++++++++++++++++++++++++++++++++++++++++++++++++++++++++++++++++++++++++
%+++++++++++++++++++++++++++++++++++++++++++++++++++++++++++++++++++++++++++
%+++++++++++++++++++++++++++++++++++++++++++++++++++++++++++++++++++++++++++

The electric form factor $D_L(p^2;T)$ and the magnetic form factor $D_T(p^2;T)$ have different behaviors
as one crosses the confinement-deconfinement phase transition temperature $T_c \approx 270$ MeV. 
For $T< T_c$, $D_L(p^2;T)$ increases with $T$ and is larger than $D_T(p^2;T)$. For $T > T_c$
$D_L(p^2;T)$ decreases with $T$ and becomes of the same order of magnitude of $D_T(p^2;T)$ 
for $T \sim 400$ MeV. The exact value of the ratio $D_L(p^2;T)/D_T(p^2;T)$ depends, for each $T$,
on the value of $p^2$. The magnetic form factor $D_T(p^2;T)$ is a decreasing function of $T$
for all the temperatures considered here. However, its nature changes when $T > T_c$ and one observe a turnover 
with $D_T(p^2;T)$ having a maximum just below $p \sim 500$ MeV.
Note, however, that the data in Fig. \ref{fig:3dtemp} does not exclude completely 
a temperature dependent shift of the turnover for the magnetic form factor, instead of a functional change of 
$D_T(p^2;T)$ as $T$ crosses the deconfinement phase transition. Indeed, if one assumes that
the temperature only induces a shift in the turnover of $D_T(p^2;T)$, 
then for $T < T_c$ the turnover should happen for very low momenta $p \sim$ 200 -- 300 MeV. 
Our simulation do not access momenta below $\sim$ 200 MeV, with the exception of p = 0 MeV, and
we are unable to distinguish the two scenarios.

For low momenta, we investigated the interpretation of the gluon form factors in terms of a quasi-particle Yukawa-like propagator, extracting the gluon screening mass $m(T)$ as a function of the temperature $T$. 
We continued previous lattice QCD studies of the gluon screening mass in Landau gauge.
Many years ago, in Ref.  ~\cite{Heller:1997nqa},
the screening electric and magnetic gluon masses were studied, for the first time, for the gauge
group $SU(2)$ and reaching temperatures as high as $10^4 \, T_c$. The authors found good agreement with finite 
temperature field theory. Since this impressive ~\cite{Heller:1997nqa} temperature range $2 T_c < T < 10^4 \, T_c$, 
lattice QCD studies started to focus in temperatures closer to the deconfinement region.
Ref. ~\cite{Nakamura:2003pu} studied $SU(3)$ gluodynamics in covariant gauges (mainly Feynman gauge) using stochastic gauge fixing, exploring with a greater detail temperatures closer to the transition temperature $T_c$
and going only up to  $ 16 \, T_c$. A finite value for the screening mass in the confinement region, i.e for $T <T_c$, was not measured. 
This is probably due to the method used to extract the mass, which relies on a point-to-all propagator; furthermore,
the gluon mass for SU(3) is for the Feynman gauge. The authors of ~\cite{Nakamura:2003pu} only concluded that
the confining screening mass was very large and possibly infinite. 
Recently, Ref. ~\cite {Maas:2011ez} was able to explore larger volumes and found finite screening masses both 
above and below $T_c$. In particular, the authors found a finite screening  mass $m_D(T)$ in the range 
$ 0 < T < 3 T_c$ and they covered in great detail the transition region of $T \sim T_c$. 
Notice a finite mass, for $T \sim 0$, is consistent with the recent study of Ref. ~\cite{Oliveira:2010xc}.
In the present work, we used for all temperatures $0< T < 2 T_c$ the same large lattice volume of (6.5 fm)$^3$. 
Our propagators are compatible with the ones of Ref. ~\cite {Maas:2011ez}, and thus we confirm finite gluon masses 
for all values of $T$, including the confining sector of gluodynamics. Note that the screening mass was measured in 
Refs. ~\cite{Heller:1997nqa,Maas:2011ez} looking only to the lowest momentum $p \sim 0$ of the gluon propagators. 
Here we explored the gluon propagator form factors $D_L(p^2;T)$ and $D_T(p^2;T)$ in a wider momentum region,
which allow us to explore different possible definitions of the gluon screening mass.

Whereas the infrared $D_L(p^2;T)$ can be described by quasi-particle Yukawa-like propagator for all the simulated
temperatures, the low momenta magnetic form factor $D_T(p^2;T)$ is not compatible with such a simple 
interpretation of the propagator.

The interpretation of the low momenta $D_L(p^2;T)$ in terms of a massive Yukawa-like propagator provides a definition of a temperature dependent
longitudinal gluon screening mass $m_D(T)$. Furthermore, this simple picture defines also an "overlap" of the gluon state with a massive quasi-particle boson
state associated with $Z(T)$. It follows that $m_D(T)$ and $Z(T)$ are sensitive to the confinement-deconfinement phase transition. The data for
$m_D(T)$ suggests that the deconfinement phase transition to be of first order.

%+++++++++++++++++++++++++++++++++++++++++++++++++++++++++++++++++++++++++++
%+++++++++++++++++++++++++++++++++++++++++++++++++++++++++++++++++++++++++++
%+++++++++++++++++++++++++++++++++++++++++++++++++++++++++++++++++++++++++++

We observe that $m_D(T)$ is a decreasing function of $T$ for $T < T_c$, 
and is an increasing function of $T$ above the confinement-deconfinement phase transition. 
Moreover, the gluon mass follows the expected perturbative functional dependence starting at 
temperatures as low as $T \sim 400$ MeV.

Besides the Yukawa approximation to the longitudinal propagator, we also consider the case of a  temperature 
dependence running gluon mass which is
motivated by the nonperturbative decoupling solution of the Dyson-Schwinger equations and that
is able to describe the lattice data at $T = 0$.
The functional form, which takes into account logarithmic corrections as suggested by perturbative QCD, is able to describe quite well both the
longitudinal and transverse gluon form factors for all the temperatures and over the full range of momenta. 
Moreover, it allows a definition of infrared mass scales $m^{IR}$ associated with $D_L(p^2;T)$ and $D_T(p^2;T)$. In 
what concerns the longitudinal
form factor $D_L(p^2;T)$ the corresponding $m^{IR}(T)$ reproduces, with large statistical errors, the behavior of the 
gluon mass taken from the Yukawa fit.
On the other hand, the magnetic $m^{IR}(T)$ is compatible with a linear behavior with $T$ over the full range of 
temperatures simulated. This linear behavior
above $T_c$ is in good agreement with the perturbative approach to hot QCD.

Last but not least, we consider a mass scale taken directly from the longitudinal gluon propagator data at zero momentum $1 / \sqrt{D_L(0)}$ as
investigated in~\cite{Maas:2011ez}. In this way, we avoid modeling the propagator. This mass scale reproduces the same pattern as observed for
the mass taken from the Yukawa or the running mass fits. Any of the definitions can be used as order parameter to identify the confinement-deconfinement
phase transition.

As an outlook of our research, with more computer power, it would be useful to further increase the lattice volume 
while decreasing the lattice spacing, in order to be able to extrapolate to the 
continuum limit. It will also be relevant to scan in more detail the transition region which would open the
possibility to investigate a possible splitting of the gluon 
mass~\cite{Pisarski:2006hz,Dumitru:2010mj,Dumitru:2012fw}.

\begin{acknowledgements}

The authors acknowledge the Laboratory for Advanced Computing at University of Coimbra for providing HPC
computing resources that have contributed to the research results reported within this paper 
(URL http://www.lca.uc.pt). 
P. J. Silva ack\-now\-led\-ges support by F.C.T. under con\-tract SFRH/BPD/40998/\-2007. 
This work was sup\-por\-ted by projects 
CERN/FP/123612/2011, CERN/FP/123620/2011 and PTDC/FIS/100968/2008, projects developed under initiative QREN financed by UE/FEDER through 
Programme COMPETE.
We thank the authors of \cite{Maas:2011ez} for sending us the data reported in 
Fig. ~\ref{fig:screeningmass_compare}.
\end{acknowledgements}

%%%%%%%%%%%%%%%%%%%%%%%%%   Bibliography   %%%%%%%%%%%
%.......................................................................................................................................
%.......................................................................................................................................
%.......................................................................................................................................
%.......................................................................................................................................
\bibliographystyle{apsrev4-1}
%\bibliography{gluonmassofT_V5}
\bibliography{gluonmassofT}

\end{document}